\documentclass[a4paper,11pt]{article}
\usepackage[margin=2.5cm]{geometry}
\usepackage{hyperref}
\usepackage{amsmath,amsfonts,amssymb,array,graphicx}
\usepackage[table]{xcolor}

\graphicspath{{fig/}}
\numberwithin{equation}{section}

\newcommand{\aver}[1]{\langle {#1} \rangle}
\newcommand{\aaver}[1]{\langle\!\langle {#1} \rangle\!\rangle}

\newcommand{\ket}[1]{| {#1} \rangle}

\newcommand{\mathfig}[1]{\raisebox{-0.5\height}{#1}}
\newcommand{\smallarray}[2]{\begin{array}{c} \scriptstyle #1 \\[-4pt] \scriptstyle #2 \end{array}}
\newcommand{\smallarrayy}[3]{\begin{array}{c} \scriptstyle #1 \\[-4pt] \scriptstyle #2 \\[-4pt] \scriptstyle #3 \end{array}}
\newcommand{\face}[1]{\mathop{\mathfig{\includegraphics{#1}}}}

\newcommand{\ds}{\displaystyle}
\newcommand{\nn}{\nonumber}
\newcommand{\wh}{\widehat}
\newcommand{\wt}{\widetilde}

\newcommand{\eproof}{{\hfill \rule{0.5em}{0.5em}}}
\newcommand{\C}{\mathcal{C}}
\newcommand{\I}{\mathcal{I}}
\newcommand{\Fc}{\mathcal{F}}
\newcommand{\V}{\mathcal{V}}
\newcommand{\Vh}{\widehat{\cal V}}
\newcommand{\Kh}{\widehat{K}}
\newcommand{\Nh}{\widehat{N}}
\newcommand{\TL}{\mathsf{TL}}
\newcommand{\id}{\mathbf{1}}
\newcommand{\Tr}{\mathrm{tr}}
\newcommand{\Cbb}{\mathbb{C}}
\newcommand{\Rbb}{\mathbb{R}}
\newcommand{\Zbb}{\mathbb{Z}}

\newcommand{\eps}{\varepsilon}
\renewcommand{\Re}{\mathrm{Re}}
\renewcommand{\Im}{\mathrm{Im}}

\newcommand{\etab}{\bar\eta}
\newcommand{\omegab}{\bar\omega}
\newcommand{\phib}{\bar\phi}
\newcommand{\psib}{\bar\psi}
\newcommand{\chib}{\bar\chi}
\newcommand{\ab}{\bar{a}}
\newcommand{\bb}{\bar{b}}
\newcommand{\cb}{\bar{c}}
\newcommand{\Fb}{\overline{F}}
\newcommand{\Lb}{\bar{L}}
\newcommand{\pb}{\bar{p}}
\newcommand{\Rb}{\overline{R}}
\newcommand{\sbar}{\bar{s}}
\newcommand{\tb}{\bar{t}}
\newcommand{\ub}{\bar{u}}
\newcommand{\vb}{\bar{v}}
\newcommand{\xb}{\bar{x}}
\newcommand{\Yb}{\overline{Y}}

\title{Non-invertible symmetries and modular invariance \\
  in lattice models}
\author{Yacine Ikhlef \\ Sorbonne Universit\'e, CNRS \\ Laboratoire de Physique Th\'eorique et Hautes \'Energies \\ LPTHE, F-75005 Paris, France}

\date{\today}

\begin{document}

\maketitle

\begin{abstract}
  We consider classical 2d lattice models with face interactions defined in terms of a fusion category. The symmetries of such models typically include an algebra of topological operators sitting on a closed path in the lattice. In the case when the face interactions obey the Temperley--Lieb (TL) relations, we present a generic algorithm to determine the decomposition of the transfer-matrix space of states as a direct sum of simple TL modules. We apply this approach to several examples, and analyse the action of topological operators. As an application, we compute the modular transformation of the irreducible TL characters at primitive roots of unity.
\end{abstract}

\section{Introduction}
\label{sec:intro}

In any discrete or continuous model of 2d Statistical Mechanics, a topological operator is defined by the insertion of modified conditions for the local degrees of freedom along a given contour, so that the resulting partition function only depends on the homotopy class of this contour. This was first studied systematically for rational Conformal Field Theories (CFTs) in the seminal works of Petkova and Zuber \cite{PZ01,Manyfaces01}, where the algebra of topological operators was derived in terms of the modular $S$-matrix of conformal characters, and it was related to the Ocneanu algebra \cite{Ocneanu99,Ocneanu00}. The lattice analogs of these operators were constructed \cite{Chui01,Chui03} in the corresponding integrable face models, using specific ``braid limits'' of the commuting transfer matrices. The results of \cite{PZ01,Manyfaces01} also found many applications to Topological Quantum Field Theories, especially after the work of \cite{Frohlich04,Frohlich07}. 
\medskip

A new perspective on the symmetries of lattice models was introduced with the concept of anyon chains associated to fusion categories \cite{Kitaev06}. The latter consist in 1d quantum systems where the quantum states of the chain live in a space of homomorphisms of the category, rather than a tensor product of spin representations as in usual quantum spin chains. The prototype of this construction is the ``golden chain'' \cite{Feiguin06}, but it equally applies to 2d statistical models \cite{Aasen16,Aasen20}. A lot of progress occurred in the understanding and classification of symmetries and dualities of these models, using the tensor-network formalism -- see \cite{Lootens23,Lootens24,Bottini26} for recent advances.
\medskip

One of the major advantages of the approach based on fusion categories is that it provides a natural construction of the algebra of topological operators and their structure constants, directly in terms of the basic data of the underlying fusion category, namely the $F$-symbols.
A particularly tractable case is when the interaction of the lattice model is given by the projector of two neighbouring ``anyons'' onto the identity object: in this case, one can show that the local operators which encode this interaction satisfy the relations of the Temperley--Lieb (TL) algebra \cite{TL71}. It turns out that many exactly solved models fit in this framework, such as the Ising, three-state Potts, Restricted Solid-On-Solid \cite{ABF84} and ADE \cite{PasquierADE87,PasquierOpContent87} lattice models. In some of these cases, it is possible to construct part of the algebra of topological operators using the braid limit of the TL transfer matrices \cite{Belletete20,Grans24}.
\medskip

In the present work, we consider a generic 2d statistical model with configurations based on a fusion category $\C$, and face interactions obeying the TL algebra. We describe the space of states of the transfer matrix, with open or closed boundary conditions, and we show that it forms respectively a module over the ordinary or affine TL category, in the sense of \cite{GL98}. Then, we present a recursive algorithm to calculate analytically the decomposition of the space of states as a direct sum of simple TL modules. This algorithm is largely inspired from previous work \cite{IMD-RSOS} on the ADE lattice models, and it is based on \textit{seed states}. The latter are states of fixed size, which generate a simple TL module when acted upon by the TL operators. For open boundary conditions, the decomposition takes the form \eqref{eq:decomp.Vbc}, and the multiplicities are encoded in a sequence of matrices $n_k$ given by Chebyshev polynomials of the defining adjacency matrix, analogously to the ``fused adjacency matrices'' introduced by \cite{BPZ98,BPPZ98} in the context of boundary CFT. For closed boundary conditions, the algorithm is similar, except that the decomposition \eqref{eq:decomp.VmK} relies on the spectrum of the cyclic translation operator $\Omega$, and the loop operator $f$.
\medskip

We then apply the above approach to three examples of critical lattice models: the Fibonacci, Ising and three-state Potts models. In each example, we write the decomposition of the space of states for fixed, periodic and twisted periodic boundary conditions, and we obtain the action of the topological operators on each term of the decomposition. In these three cases, we observe that the algebra of TL endomorphisms of the space of states is generated by the topological operators associated to the simple objects and the automorphisms of $\C$. Moreover, we discuss the scaling limit in terms of bulk CFT. Besides these examples, we consider a non-critical model based on the psu(2)$_5$ fusion rules (see \cite{Blakeney26}), where the space of states takes the form of an infinite direct sum of simple TL modules.
\medskip

Finally, similarly to the Virasoro modules, we define modular characters as traces on the affine TL simple modules. A natural question is then to determine how these characters transform under the modular group. To answer this question when the loop weight is of the form $\beta=2\cos(\pi/h)$ where $h$ is an integer, we first analyse the unitary RSOS models of \cite{ABF84} with twisted periodic boundary conditions, and determine the decomposition of the space of states and the eigenvalues of the topological operators. By reversing the logic of \cite{PZ01}, we use these results to derive the modular transformation of the irreducible TL characters.

\section{Lattice models with local Temperley-Lieb interactions}
\label{sec:models}

\subsection{Definition of the models}
\label{sec:def}

\paragraph{Face model.}
A \textit{face model} on the square lattice is a statistical model, defined in terms of a fusion category $\C$,
and a simple object $a$ of $\C$ such that $\ab=a$.
Let $G=\Nh(a)$ be the corresponding adjacency matrix.
Each site $i$ of the lattice carries a simple object $x_i$, so that $G_{x_i,x_j}=1$ for any pair of adjacent sites $(i,j)$ on the square lattice. The Boltzmann weight of such a \textit{spin configuration} is the product of face weights, given by
\begin{equation} \label{eq:face}
  W_v\left(
  \sideset{_{\ds w}^{\ds z}}{_{\ds x}^{\ds y}}{\face{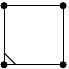}}
  \right) 
  = \delta_{wy} \left[\frac{d(x)d(z)}{d(w)d(y)}\right]^{1/4}
  + v \, \delta_{xz}\, \left[\frac{d(w)d(y)}{d(x)d(z)}\right]^{1/4} \,,
\end{equation}
where $d(x)$ denotes the quantum dimension, and $v>0$ is a local coupling constant.
Consider the partition function on an $N \times M$ lattice, say with fixed boundary conditions.
\begin{center}
  \includegraphics{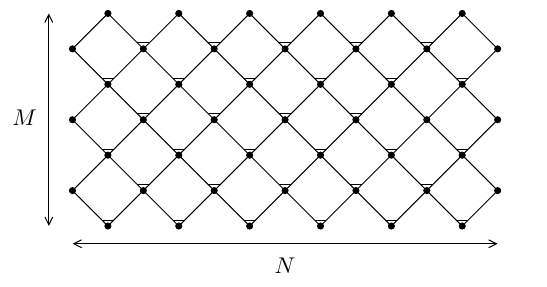}
\end{center}
Each horizontal slice of the lattice carries a sequence of simple objects
$$\includegraphics{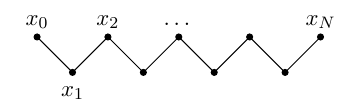}$$
such that and $G_{x_i,x_{i+1}}=1$ for any $i=0,\dots,N-1$. Such sequences are denoted $[x_0,x_1,\dots,x_N]$, and they label the basis states in the transfer-matrix formalism.

\paragraph{Fixed boundary conditions.}
Given a pair of simple objects $(b,c)$, one fixes the left and right spins to $b$ and $c$ respectively. The space of states is then
\begin{equation}
  \V_{bc}(N) := \mathrm{span} \Big(
  [x_0,x_1,\dots,x_N] \,, \quad x_0=b \,, \quad x_N=c
  \Big) \,.
\end{equation}
The overall Boltzmann weight of a spin configuration is invariant under the ``change of gauge''
\begin{equation}
  W_v\left(\sideset{_{\ds w}^{\ds z}}{_{\ds x}^{\ds y}}{\face{face.pdf}} \right)
  \to \left[\frac{d(w)d(y)}{d(x)d(z)}\right]^{1/4} \,W_v\left(\sideset{_{\ds w}^{\ds z}}{_{\ds x}^{\ds y}}{\face{face.pdf}} \right) \,.
\end{equation}
Upon this modification, a face at position $j$ in the transfer matrix corresponds to the action of a face operator of the form $(\id+v e_j)$, where
\begin{equation} \label{eq:ej.[x]}
  e_j \cdot [x_0,x_1,\dots,x_N]
  := \sum_{x'_j=1}^n \delta_{x_{j-1},x_{j+1}} \, G_{x_{j-1},x'_j}
  \, \sqrt{\frac{d(x'_j)d(x_j)}{d(x_{j-1})d(x_{j+1})}}
  \, [x_0\dots x_{j-1},x'_j,x_{j+1} \dots x_N] \,,
\end{equation}
where the integers $1,\dots,n$ label the simple objects of the category $\C$.
The operators  $e_1,\dots,e_{N-1}$ satisfy the relations
\begin{equation} \label{eq:TL.relations.ej}
  e_j^2 = \beta\, e_j \,,
  \qquad e_j e_{j\pm 1} e_j = e_j \,,
  \qquad e_j e_k=e_k e_j \quad \text{if } |j-k|>1 \,,
\end{equation}
with the loop weight $\beta=d(a)$. The abstract algebra generated by $e_1,\dots,e_{N-1}$ subject to these relations is called the Temperley--Lieb algebra with loop weight $\beta$, and we denote it by $\TL(N)$.
It is particularly useful to view $\V_{bc}(N)$ as the morphism space from $b$ to $(a^{\otimes N}\otimes c)$ in the category~$\C$:
\begin{equation}
  \V_{bc}(N) = V_b^{a^Nc} \,,
  \qquad [x_0,\dots,x_N] = \mathfig{\includegraphics{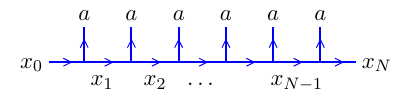}}
\end{equation}
with $x_0=b$ and $x_N=c$.
In this language, the operator \eqref{eq:ej.[x]}  can be written $e_j=c^\dag_jc_j^{\phantom\dag}$, where $c_j:\V_{bc}(N)\to \V_{bc}(N-2)$ and $c^\dag_j:\V_{bc}(N)\to \V_{bc}(N+2)$ are defined as
\begin{equation} \label{eq:cj.u}
  c_j \cdot u = \raisebox{-5pt}{\includegraphics{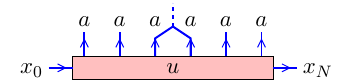}}  
  \qquad
  c^\dag_j \cdot u = \raisebox{-5pt}{\includegraphics{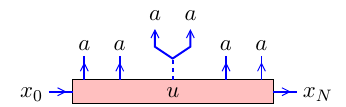}} 
\end{equation}
with $x_0=b$ and $x_N=c$.
Here the dotted line represents the identity object $1$.
The action of $c_j$ and $c_j^\dag$ on the basis states of $\V_{bc}$ reads
\begin{equation} \label{eq:cj.[x]}
  \begin{aligned}
    & c_j \cdot [x_0,\dots,x_N] = \delta_{x_{j-1},x_{j+1}}\, \sqrt{\frac{d(a)d(x_j)}{d(x_{j+1})}} \, [x_0\dots x_{j-1},x_{j+2}\dots x_N] \,, \\
    & c^\dag_j \cdot [x_0,\dots,x_N] = \sum_{x'_j} G_{x_{j-1},x'_j} \, \sqrt{\frac{d(x'_j)}{d(a)d(x_{j-1})}}
    \, [x_0\dots x_{j-1},x'_j,x_{j-1}\dots x_N] \,.
  \end{aligned}
\end{equation}
\medskip

Note that the operators $e_j$ (as well as $c_j$ and $c_j^\dag$) are defined completely in terms of the adjacency matrix $G=\Nh(a)$. Indeed, the quantum dimensions $d(x)$ appearing in \eqref{eq:ej.[x]} and \eqref{eq:cj.[x]} are the coefficients of the dominant eigenvector of $G$.

\paragraph{Twisted periodic boundary conditions.}
We consider twisted boundary conditions labelled by a pair $(m,K)$, where $m$ is a simple object of $\C$, and $K$ is an automorphism of $\C$ as in \eqref{eq:automorphism}, such that $K(a)=a$. We thus define the space of states
\begin{equation}
  \Vh_{m,K}(N):= \mathrm{span} \Big(
  [x_0,x_1,\dots,x_N] \,, \quad N_{K(x_0),m}^{x_N}=1
  \Big) \,.
\end{equation}
In terms of morphisms of $\C$, we have
\begin{equation}
  \begin{aligned}
    & \Vh_{m,K}(N)= \bigoplus_{x=1}^n V_{x}^{a^N K(x) m} \,, \\
    & [x_0,x_1,\dots,x_N] = \mathfig{\includegraphics{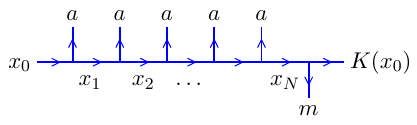}} 
  \end{aligned}
\end{equation}
The operators $e_1,\dots,e_{N-1}$ are defined on $\Vh_{m,K}(N)$ as in \eqref{eq:ej.[x]}.
Additionally, for $N \geq 1$, we define the shift operators
\begin{equation} \label{eq:Omega.u}
  \begin{aligned}
    &\Omega \cdot u := \sum_{x'=1}^n
    \sqrt{\frac{d(x')}{d(x)d(a)}} \quad \mathfig{\includegraphics{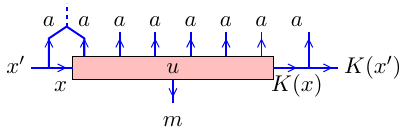}} \quad, \\
    &\Omega^\dag \cdot u := \sum_{x'=1}^n
    \sqrt{\frac{d(x')}{d(x)d(a)}} \quad \mathfig{\includegraphics{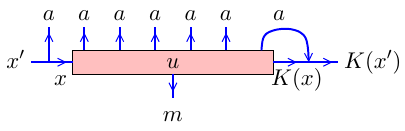}} \quad,
  \end{aligned}
\end{equation}
for any $u\in V_x^{a^NK(x)m}$.
For $N=0$, instead of $\Omega$ and $\Omega^\dag$, we introduce the loop operator
$f=f_a$, with
\begin{equation} \label{eq:f.u}
  f_a \cdot u :=\sum_{x'=1}^n
    \sqrt{\frac{d(x')}{d(x)d(a)}} \quad \mathfig{\includegraphics{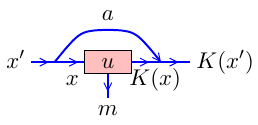}} \quad\qquad,
\end{equation}
for any $u\in V_x^{K(x)m}$.
The action of $\Omega,\Omega^\dag,f_a$ on the basis states reads
\begin{equation} \label{eq:Omega.x}
  \begin{aligned}
    &\Omega \cdot [x_0,x_1,\dots,x_N] =
    \sum_{y=1}^n \left(F_{x_N}^{aK(x_1)m}\right)_{yK(x_0)} \, [x_1,x_2,\dots,x_N,y] \,, \\
    &\Omega^\dag \cdot [x_0,x_1,\dots,x_N] =
    \sum_{y=1}^n \left(F_{x_{N-1}}^{aK(x_0)m}\right)^*_{x_NK(y)} \, [y,x_0,x_1,\dots,x_{N-1}] \,, \\
    & f_a\cdot [x] = \sum_{y=1}^n \left(F_x^{aK(y)m}\right)_{yK(x)} \, [y] \,,
  \end{aligned}
\end{equation}
where $F_w^{xyz}$ denotes the $F$-symbol of the fusion category $\C$.
\medskip

For $N\geq 2$, we consider $e_1,\dots,e_{N-1}$, together with the shift operators $\Omega,\Omega^\dag$, and $e_N=e_0:=\Omega e_1 \Omega^\dag$. These operators satisfy the relations
\begin{equation}
  \begin{aligned}
    & e_j^2 = \beta \, e_j \,,
    &\quad& e_je_{j\pm 1} e_j = e_j \,,
    &\quad& e_je_k=e_ke_j \quad \text{if }|j-k|>1 \\
    & \Omega\Omega^\dag = \Omega^\dag\Omega = \id \,,
    && e_j\Omega = \Omega e_{j+1} \,,
    && \Omega^2 e_1 = e_{N-1}e_{N-2}\dots e_1 \,,
  \end{aligned}
\end{equation}
where the indices $j,k$ are considered modulo $N$, and $\beta=d(a)$.
The abstract algebra generated by $e_1,\Omega,\Omega^\dag$ is called the affine Temperley--Lieb algebra with loop weight $\beta$, and we denote it as $\TL^a(N)$.
For $N=1$, the algebra $\TL^a(1)$ is generated by $\Omega,\Omega^\dag$ subject to $\Omega\Omega^\dag = \Omega^\dag\Omega = \id$. For $N=0$, one defines $\TL^a(0)=\mathrm{span}(f^n, n\in \Zbb_{\geq 0})$.
\medskip

The operators $e_1,\dots,e_{N-1}$ acting on $\Vh_{m,K}(N)$ decompose as $e_j=c_j^\dag c_j^{\phantom\dag}$, with $c_j$ and $c^\dag_j$ given in \eqref{eq:cj.u}. Additionally, we define $e_0:=c_0^\dag c_0^{\phantom\dag}$, where $c_0:=c_1\Omega$ and $c_0^\dag:=\Omega c_1^\dag$. As a result, $e_1,\dots,e_{N-1}$ are defined completely in terms of the adjacency matrix $G=\Nh(a)$, whereas $\Omega$, $\Omega^\dag$ and $e_0$ (or $f$ if $N=0$) include some coefficients which depend on the $F$-symbols of $\C$.
\medskip

Note that in the case $(m,K)=(1,\id)$, one recovers the periodic boundary conditions $x_N=x_0$. We denote $\Vh(N):=\Vh_{1,\id}(N)$ the corresponding space. The shift operator $\Omega$ acting on $\Vh(N)$ is then the usual left cyclic translation operator
\begin{equation}
  \Omega \cdot [x_0,x_1,\dots,x_N] = [x_1,x_2,\dots,x_N,x_1] \,,
  \qquad \Omega^N=\id \,,
\end{equation}
whereas $f_a$ acts as $\Nh(a)$ on $\Vh(0)$. However, for a general twisted space $\Vh_{m,K}(N)$, the operator $\Omega^N$ remains central (i.e. it commutes with $e_0,\dots,e_N$) but it is not necessarily a multiple of $\id$.
\medskip

The face operator at position $j=0$ is given by $\Omega(\id+ v e_1)\Omega^\dag$, and thus the twisted periodic boundary conditions in the horizontal direction correspond to the insertion of a ``seam'' with modified Boltzmann weights as in the following picture:
\begin{center}
  \includegraphics{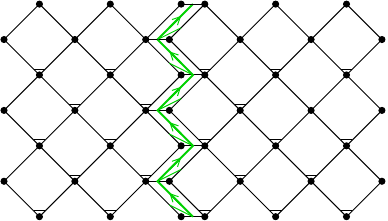}
\end{center}
The elementary weights of the seam, encoded in \eqref{eq:Omega.x}, are given by
\begin{equation} \label{eq:seam}
  \begin{aligned}
    & \sideset{_{\ds x}^{\ds x'}}{_{\ds y}^{\ds y'}}{\face{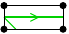}}
    = \Fc_{[x'y'],[xy]}^{(m,K)} = \left(F_{x'}^{aK(y)m}\right)_{y',K(x)} \,, \\
    & \sideset{_{\ds x}^{\ds x'}}{_{\ds y}^{\ds y'}}{\face{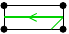}}
    = \overline{\Fc}_{[x'y'],[xy]}^{(m,K)} = \left(F_x^{aK(y')m}\right)^*_{y,K(x')} \,.
  \end{aligned}
\end{equation}

\subsection{Topological symmetries}
\label{sec:topo}

In analogy with the concepts introduced in \cite{PZ01} for CFT models, we define lattice topological operators as follows. Let $A,B$ be two $\TL$ modules. A topological operator $X:A\to B$ is a collection\footnote{In the following, to lighten the notation, we shall drop the index $N$, and write $X$ instead of $X_N$.} of linear maps $X_N$ from $A(N)$ to $B(N)$, such that $X_N\cdot \lambda = \lambda\cdot X_{N'}$ for any $\lambda\in\TL(N',N)$. In other words, a topological operator is a morphism $A\to B$ of $\TL$ modules (see Appendix~\ref{app:TL}). Hence, in particular, we shall denote by $\mathrm{End}(A)$ the algebra of topological operators from $A$ to itself. The same definitions apply to topological operators on $\TL^a$ modules.

In this section, we review the construction of families of topological operators on the modules $\V_{bc}$ (resp. $\Vh_{m,K}$), whose action corresponds to modifying the Boltzmann weights along a path connecting the two boundaries (resp. along a non-trivial closed path). The fact that $X$ commutes with the generating algebra $\TL$ or $\TL^a$ ensures that $X$ depends only on the homotopy class of the defining path.

\paragraph{Periodic boundary conditions.}
Let us first recall the construction \cite{Feiguin06} of topological operators $Y_p$ indexed by a simple object $p$ of $\C$, acting on the space $\Vh(N)$ with periodic boundary conditions.
One defines
\begin{equation}
  Y_p \cdot u := \sum_{x'=1}^n \sqrt{\frac{d(x')}{d(x)d(p)}}
  \quad \mathfig{\includegraphics{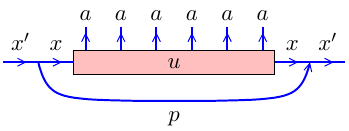}}
\end{equation}
for any $u \in V_x^{a^Nx}$.
By construction, $Y_p$ commutes with \eqref{eq:cj.u}. Using appropriate $F$-moves \eqref{eq:F-move}, one shows easily that $Y_p$ commutes also with $\Omega,\Omega^\dag,f$ defined in \eqref{eq:Omega.u} and \eqref{eq:f.u}. Hence, $Y_p$ is topological, i.e. it commutes with $\TL^a(N,N')$.
The action of $Y_p$ on the basis states of $\Vh(N)$ reads
\begin{equation}
  Y_p \cdot [x_0,x_1,\dots,x_N]
  = \sum_{[x'_0,x'_1,\dots,x'_N]} \Fc_{\boldsymbol{x'},\boldsymbol{x}}^{(p)} \,
  [x'_0,x'_1,\dots,x'_N] \,,
\end{equation}
where
\begin{equation}
  \Fc_{\boldsymbol{x'},\boldsymbol{x}}^{(p)} := \prod_{j=0}^{N-1}
  \Fc_{[x'_j,x'_{j+1}],[x_j,x_{j+1}]}^{(p,\id)} \,,
\end{equation}
and $\Fc_{[x',y'],[x,y]}^{(p,\id)}$ is given in \eqref{eq:seam}.
Hence, the action of $Y_p$ corresponds to the insertion of a horizontal seam with the same weights as in the vertical seam encoding the twisted boundary conditions of $\Vh_{p,\id}$. For $p=1$, one has $Y_1=\id$.
\medskip

Additionally, to any $\C$-automorphism $L$ such that $L(a)=a$, we associate the operator $Q_L$ acting on $\Vh(N)$ as
\begin{equation} \label{eq:QL.x}
  Q_L \cdot [x_0,\dots,x_N] := [L(x_0),\dots,L(x_N)] \,.
\end{equation}
Using the action on the basis states, one easily shows that $Q_L$ commutes with $\TL^a(N,N')$.
\medskip

The topological operators of type $Y_p$ and $Q_L$ obey the composition rules
\begin{equation}
  Y_p \cdot Y_{p'} = \sum_{p''=1}^n N_{pp'}^{p''} \, Y_{p''} \,,
  \qquad Q_L \cdot Q_{L'} = Q_{LL'} \,,
  \qquad Q_L\cdot Y_p= Y_{L(p)}\cdot Q_L \,,
\end{equation}
where $N_{pp'}^{p''}$ are the fusion numbers of the simple objects of $\C$.

\paragraph{Twisted periodic boundary conditions.}
For any pair of simple objects $(m,m')$ and any automorphism $K$, one has the family of topological operators (also called lasso maps) \cite{Verstraete17}
\begin{equation}
  Y_{p,\alpha}:\Vh_{m,K}(N) \to \Vh_{m',K}(N) \,,
\end{equation}
indexed by two simple objects $p,\alpha$.
The operator $Y_{p,\alpha}$ is defined as
\begin{equation} \label{eq:Ypa}
  Y_{p,\alpha} \cdot u :=
  \sum_{x'=1}^n \sqrt{\frac{d(x')}{d(x)d(p)}}
  \quad \mathfig{\includegraphics{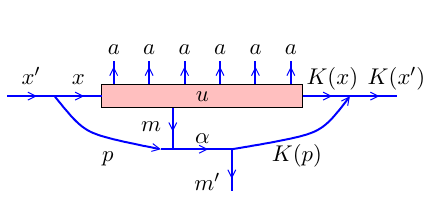}} 
\end{equation}
for any $u \in V_x^{a^NK(x)m}$.
The action of $Y_{p,\alpha}$ on the basis states of $\Vh_{m,K}(N)$ reads
\begin{equation}
  Y_{p,\alpha} \cdot [x_0,x_1,\dots,x_N]
  = \sum_{[x'_0,x'_1,\dots,x'_N]} \Fc_{\boldsymbol{x'},\boldsymbol{x}}^{(p)} \,
  \, U_{[x'_N,x'_0],[x_N,x_0]}^{(p,\alpha)} \, [x'_0,x'_1,\dots,x'_N] \,,
\end{equation}
where the sum is on basis states of $\Vh_{m',K}(N)$, and
\begin{equation}
  U_{[x',y'],[x,y]}^{(p,\alpha)}:= \sqrt{\frac{d(m)d(p)}{d(\alpha)}}
  \, \left(F_{x'}^{K(y)K(p)m'} \right)_{\alpha K(y')}^*
  \, \left(F_{x'}^{K(y)mp} \right)_{\alpha x} \,.
\end{equation}
\medskip

For any automorphism $L$ such that $L(a)=a$, one defines the operator $Q_L:\Vh_{m,K}(N) \to \Vh_{L(m),LKL^{-1}}(N)$ as in \eqref{eq:QL.x}.
\medskip

The operators of type $Y_{p,\alpha}$ and $Q_L$ obey the composition rules
\begin{equation}
  \begin{aligned}
    & Y_{p',\alpha'} \cdot Y_{p,\alpha}
    = \sum_{p',\alpha''} \sqrt{\frac{d(p)d(p')}{d(p'')}}
    \, \left(F_\alpha^{K(p)\alpha'\pb'} \right)_{m'\alpha''}^*
    \left(F_m^{\alpha''\pb'\pb} \right)_{\pb''\alpha}
    \left(F_{\alpha''}^{K(p)K(p')m''} \right)^*_{\alpha'K(p'')} \, Y_{p'',\alpha''} \,, \\
    & Q_L \cdot Y_{p,\alpha} = Y_{L(p),L(\alpha)} \cdot Q_L \,, 
    \qquad Q_L \cdot Q_{L'} = Q_{LL'} \,.
  \end{aligned}
\end{equation}

The above relations define the tube algebra, whose representation theory involves some elaborate tools from category theory (see \cite{Verstraete17}). In the particular case when $\C$ is a \textit{braided} fusion category, the tube algebra admits two commuting subalgebras which are both equivalent to the fusion algebra of $\C$.
Indeed, consider the topological operators acting in $\Vh_{m,\id}(N)$:
\begin{equation}
  Y_p := \sum_{\alpha=1}^n \sqrt{\frac{d(\alpha)}{d(m)d(p)}} \, R_\alpha^{mp} \, Y_{p,\alpha} \,,
  \qquad \Yb_p:= \sum_{\alpha=1}^n \sqrt{\frac{d(\alpha)}{d(m)d(p)}} \, \Rb_\alpha^{mp} \, Y_{p,\alpha} \,,
\end{equation}
where $R_\alpha^{mp}$ and $\Rb_\alpha^{mp}$ are the coefficients of the braid matrices of $\C$.
Pictorially, these operators are represented as
\begin{equation} \label{eq:Yp.braid}
  \begin{aligned}
    & Y_p \cdot u := \sum_{x'=1}^n \sqrt{\frac{d(x')}{d(x)d(p)}}
    \quad \mathfig{\includegraphics{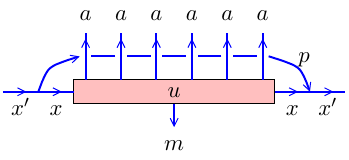}} \\
    & \Yb_{\!\!p} \cdot u := \sum_{x'=1}^n \sqrt{\frac{d(x')}{d(x)d(p)}}
    \quad \mathfig{\includegraphics{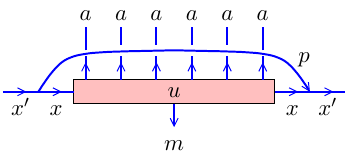}}
  \end{aligned}
\end{equation}
for any $u \in V_x^{a^Nx}$, where we have used the braid relations to move the line with label $p$ to the top of the diagrams.
For $p=1$, we have $Y_1=\Yb_{\!\!1}=\id$.
The braid relations also yield the composition rules
\begin{equation}
  Y_p \cdot Y_{p'} = \sum_{p''=1}^n N_{pp'}^{p''} \, Y_{p''} \,,
  \qquad \Yb_{\!\!p} \cdot \Yb_{\!\!p'} = \sum_{p''=1}^n N_{pp'}^{p''} \, \Yb_{\!\!p''} \,,
  \qquad Y_p \Yb_{\!\!p'} = \Yb_{\!\!p'} Y_p \,.
\end{equation}
If one assumes moreover that the fusion rules are such that $a\otimes a = \id \oplus a'$, where $a'$ is a simple object, then one can show that the matrices $R_{aa}$ and $\Rb_{aa}$ are of the form
\begin{equation} \label{eq:Raa}
  R_{aa} = y \,\id_{aa} + y^{-1} \,\psi_1^{aa}\circ \psi_{aa}^1 \,,
  \qquad \Rb_{aa} = y ^{-1}\,\id_{aa} + y \,\psi_1^{aa}\circ \psi_{aa}^1 \,,
\end{equation}
where $y\in \Cbb^\times$ is such that $y^2=q$ or $y^2=q^{-1}$.
In this case, the operators $Y_a,\Yb_{\!\!a}$ coincide respectively with the $\TL^a$ braid transfer matrices $F,\Fb$ \eqref{eq:F} acting on $\Vh_{m,\id}(N)$, and hence they have eigenvalues given by \eqref{eq:F.I} on the simple submodules of $\Vh_{m,\id}$.

\paragraph{Fixed boundary conditions.}
Similarly to the case of periodic boundary conditions, we define for any simple object $p$, the operator $Y_p:\V_{bc}(N)\to\V_{b'c'}(N)$ as
\begin{equation}
  Y_p \cdot u :=
  \mathfig{\includegraphics{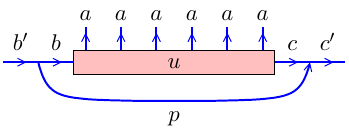}}
\end{equation}
which yields a topological operator by construction.
The action of $Y_p$ on the basis states of $\V_{bc}(N)$ reads
\begin{equation}
  Y_p \cdot [x_0,\dots,x_N] = \sqrt{\frac{d(c)d(p)}{d(c')}}
  \sum_{x'_1,\dots,x'_{N-1}} \Fc_{\boldsymbol{x'},\boldsymbol{x}}^{(p)} \, [b',x'_1,\dots,x'_{N-1},c'] \,,
\end{equation}
where $(x_0,x_N)=(b,c)$ and $(x'_0,x'_N)=(b',c')$.
The composition rules are
\begin{equation}
  Y_{p'} \cdot Y_p = \sum_{p''} \sqrt{\frac{d(p)d(p')}{d(p'')}}
  \left(F_{b''}^{bpp'} \right)_{p''b'} \, \left(F_{c''}^{cpp'}\right)_{c'p''}^* \, Y_{p''} \,.
\end{equation}
\medskip

For any automorphism $L$ such that $L(a)=a$, one defines the operator $Q_L:\V_{bc}(N) \to \V_{L(b)L(c)}(N)$ as in \eqref{eq:QL.x}.
\medskip

Like in the case of periodic twisted boundary conditions, if $\C$ is a braided category then one can construct a subalgebra of topological operators which obey the fusion rules of $\C$.
We introduce operators acting in $\V_{bc}(N)$ as 
\begin{equation}
  X_p \cdot u = \quad \mathfig{\includegraphics{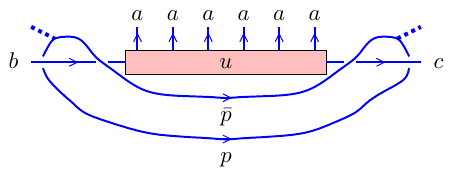}} 
\end{equation}
These operators are topological by construction. Moreover, they decompose over the operators $Y_m$ as
\begin{equation}
  X_p = \sum_m \sqrt{\frac{d(m)}{d(p)d(\pb)}} \, x_{pmb}^* \,x_{pmc} \, Y_m \,,
  \qquad \text{where} \quad x_{pmc} = \sum_r \sqrt{\frac{d(p)d(r)}{d(m)d(c)}}\, \left(F_c^{c\pb p}\right)^*_{mr} \, R_r^{\pb c}\, \Rb_r^{c\pb} \,.
\end{equation}
Using the braid relations, we obtain the commutation rules
\begin{equation}
  X_p \cdot X_{p'} = \sum_{p''} N_{pp'}^{p''} \, X_{p''} \,.
\end{equation}
Additionally, if \eqref{eq:Raa} holds, then using the braid relations, one can move the lines indexed by $p,\pb$ to the top of the diagram (similarly to the operators \eqref{eq:Yp.braid}), and thus show that $X_a$ coincides with the braid transfer matrix $G$ \eqref{eq:G}. Hence, in this case, $X_a$ has eigenvalues given by \eqref{eq:F.I} on the simple submodules of $\V_{bc}$.

\subsection{Decomposition over simple Temperley--Lieb modules}
\label{sec:decomp}

In this section, we refer to the classification of simple Temperley--Lieb modules found in \cite{GL98}: the corresponding material is reviewed in Appendix~\ref{app:TL}.
When studying the face model associated to a pair $(\C,a)$, we consider two distinct cases, depending on the value of the quantum dimension~$d(a)$.
\begin{enumerate}
\item If $0<d(a)<2$ then one can show that $d(a)$ is of the form $2\cos(\pi/h)$, where $h$ is a positive integer. We then set $q=e^{i\pi(h-1)/h}$. Moreover, we assume that all the eigenvalues of the adjacency matrix $G=\Nh(a)$ are of the form $\lambda_j(a)=2\cos(\pi m_j/h)$, with $m_j\in \{1,2,\dots,h-1\}$. The integers $m_j$ are called the exponents of $G$.
\item If $d(a) \geq 2$, then $d(a)$ is of the form $2\cosh \gamma$ with $\gamma \in \Rbb$. We set $q=-e^\gamma$, and $h=\infty$ by convention. We also parameterise the eigenvalues of $\Nh(a)$ as $\lambda_j(a)=2\cos(\pi \theta_j)$, with $\theta_j \in \Cbb$.
\end{enumerate}
In both cases, we have $d(a)=-q-q^{-1}$.

\paragraph{Fixed boundary conditions.}
Consider the space of states $\V_{bc}(N)$ with fixed boundary conditions, in the face model defined by $(\C,a)$.
The spaces $\V_{bc}(0),\V_{bc}(1),\V_{bc}(2)\dots$ equipped with the action of $c_j,c^\dag_j$ \eqref{eq:cj.u} form a $\TL$ module. This means that the diagrams $\lambda \in \TL(N,N')$ obtained by composing operators of the form $c_j$ and $c^\dag_j$ act from $\V_{bc}(N')$ to $\V_{bc}(N)$.
Moreover, the inner product on $\V_{bc}(N)$ defined by
\begin{equation} \label{eq:<x,y>}
  \aver{[x_0,\dots,x_N],[y_0,\dots,y_N]} := d(a)^{N/2} \, \delta_{x_0,y_0} \dots \delta_{x_N,y_N} 
\end{equation}
is invariant, namely it satisfies $\aver{c_ju,v}=\aver{u,c_j^\dag v}$ for any $u \in \V_{bc}(N),\ v\in \V_{bc}(N-2)$ and $j=1,\dots,N-1$.
\medskip

Based on these properties, let us show that the module $\V_{bc}(N)$ decomposes as
\begin{equation} \label{eq:decomp.Vbc}
  \V_{bc}(N) \equiv \bigoplus_{k=0,\tfrac{1}{2},\dots,\tfrac{h}{2}-1} (n_k)_{bc} \, I_k(N) \,.
\end{equation}
Here, $I_k$ is the simple module with $2k$ legs, and the multiplicities are given by the ``fused adjacency matrices'' \cite{BPZ98} defined  as
\begin{equation}
  n_k = \begin{cases}
    U_{2k}[\Nh(a)/2] &\text{if } k \in \{0,\tfrac{1}{2},\dots,\tfrac{h}{2}-1\} \\
    0 & \text{otherwise,}
  \end{cases}
\end{equation}
where $U_{2k}$ is the $(2k)$-th Chebyshev polynomial of the second kind (see Appendix~\ref{app:cheby}). Note that the direct sum in \eqref{eq:decomp.Vbc} is finite if $d(a)<2$, and infinite if $d(a)\geq 2$. However, in the latter case, for any value of $N$, all terms with $k>N/2$ vanish.
\medskip

The first ingredient to prove \eqref{eq:decomp.Vbc} is the equality of dimensions:
\begin{equation} \label{eq:dim.Vbc}
  \dim \V_{bc}(N) = \sum_{k=0,\tfrac{1}{2},\dots,\tfrac{h}{2}-1} (n_k)_{bc} \, \dim I_k(N) \,.
\end{equation}
In order to establish this equality, we use the orthonormal basis of eigenvectors of $\Nh(a)$, given by the normalised fusion characters
\begin{equation}
  w_j(a) := \frac{\lambda_j(a)}{|\lambda_j|} \,, \qquad
  |\lambda_j| := \sqrt{\sum_{a=1}^n |\lambda_j(a)|^2} \,.
\end{equation}
We write the LHS of \eqref{eq:dim.Vbc} as
\begin{alignat}{1}
  \dim \mathcal{V}_{bc}(N) &= \left[\wh{N}(a)^N\right]_{bc}
  = \sum_{j=1}^n w_j(b)^* w_j(c) \, \lambda_j(a)^N \nn\\
  & = \begin{cases}
    \ds \sum_{j=1}^n \sum_{k=0,\frac 12,\dots,h-\frac12} w_j(b)^* w_j(c)\, D_k(N) \,\cos \frac{2\pi km_j}{h} & \text{if } d(a)<2 \\
    \ds \sum_{j=1}^n \sum_{k\in\Zbb/2} w_j(b)^* w_j(c)\, d_k(N) \,\cos 2k\theta_j & \text{if } d(a) \geq 2
  \end{cases}
\end{alignat}
where we have used the binomial identities \eqref{eq:Newton}--\eqref{eq:Newton2}.
If $d(a)<2$, the RHS of \eqref{eq:dim.Vbc} reads
\begin{alignat}{1}
  &\sum_{k=0,\frac 12,\dots,\frac h2-1} (n_k)_{bc}\, \dim I_k(N)
  = \sum_{k=0,\frac 12,\dots,\frac h2-1}\sum_{j=1}^n w_j(b)^* w_j(c) \,  
   U_{2k}(\lambda_j(a)/2) \,[D_k(N)-D_{k+1}(N)] \nn\\
   &\quad=  \sum_{j=1}^n w_j(b)^* w_j(c) \,\Bigg[ \sum_{k=1,\frac 32\dots,\frac h2-1}
     (U_{2k}-U_{2k-2})\left(\cos\frac{\pi m_j}{h}\right) \, D_k(N)
     + D_0(N)  \nn \\
     &\qquad\qquad\qquad\qquad\qquad
     + 2\cos\frac{\pi m_j}{h} \, D_{1/2}(N) + \cos \pi m_j \, D_{h/2}(N)
     \Bigg] \nn\\
   &\quad= \sum_{j=1}^n \sum_{k=0,\frac12,\dots,h-\frac12} w_j(b)^* w_j(c) \, D_k(N)\,\cos\frac{2\pi km_j}{h}  \,.
\end{alignat}
where we have used the property \eqref{eq:diff.Um}.
If $d(a)\geq 2$, we obtain
\begin{alignat}{1}
  &\sum_{k\in\Zbb_{\geq 0}/2} (\mu_k)_{bc}\, \dim I_k(N)
  = \sum_{k\in\Zbb_{\geq 0}/2} \sum_{j=1}^n w_j(b)^* w_j(c) \, 
   U_{2k}(\lambda_j(a)/2) \,[d_k(N)-d_{k+1}(N)] \nn\\
   &\quad= \sum_{j=1}^n w_j(b)^* w_j(c) \,\left[
     \sum_{k=1,\frac32,2,\dots\infty} (U_{2k}-U_{2k-2})(\cos\theta_j) \, d_k(N)
     + d_0(N) + 2\cos\theta_j \, d_{1/2}(N)
     \right] \nn \\
   &\quad= \sum_{j=1}^n \sum_{k\in \Zbb}w_j(b)^* w_j(c) \, d_k(N)\,\cos 2k\theta_j \,.
\end{alignat}
Hence, in both cases, the formula \eqref{eq:dim.Vbc} is proved.
\medskip

The next key ingredient to prove \eqref{eq:decomp.Vbc} consists in exploiting the fact that the simple modules are generated by the action of $\TL$ diagrams on a single state, i.e.
\begin{equation}
  I_k(N) = \TL(N,2k) \cdot u_k \,,
\end{equation}
where $u_k$ is the unique basis state of $I_k(2k)=W_k(2k)$.
We say that $u_k$ is a seed state for $I_k$.
Similarly, for any $\TL$-module $A$ and any $k \in \Zbb_{\geq 0}/2$, a seed state of type $k$ in $A$ is defined as a nonzero state $\xi \in A(2k)$ such that $c_1\cdot \xi=c_2\cdot \xi=\dots = c_{2k-1}\cdot \xi=0$.
\medskip

For instance, a seed state of type $k=1$ in $\V_{bc}$ is a nonzero state of the form $\xi=\sum_x \alpha_x [b,x,c]$ in $\V_{bc}(2)$, where $\alpha_x\in\Cbb$, and such that $c_1\cdot \xi=0$. Given the expression \eqref{eq:cj.[x]} of $c_j$ in $\V_{bc}$, we have
$$c_1\cdot [b,x,c] = \delta_{bc} \, \sqrt{\frac{d(x)d(a)}{d(b)}} \, [b,c] \,.$$
Hence, if $b\neq c$ then any basis state $[b,x,c]$ of $\V_{bc}(2)$ is a seed state of type $k=1$. If $b=c$ then any nonzero solution to the linear equation $\sum_x  \alpha_x \sqrt{d(x)}=0$ yields a seed state of type $k=1$.
\medskip

In Appendix~\ref{app:TL} we show that, if the loop weight $\beta$ is real and the module $A$ is equipped with an invariant inner product~$\aver{\,,\,}$ then
\begin{equation}
  \TL(N,2k) \cdot \xi \equiv I_k(N) \,,
\end{equation}
for any seed state $\xi$ of type $k$ in $A$. Moreover, if $\xi,\xi'$ are two seed states respectively of types $k,k'$ in $A$, with  $\aver{\xi,\xi'}=0$ in the case $k=k'$, then $\TL(N,2k) \cdot \xi$ is orthogonal to $\TL(N,2k) \cdot \xi'$.
\medskip

Let us now prove \eqref{eq:decomp.Vbc} by constructing the seed states recursively.
More specifically, let us construct, for any $k=0,\tfrac 12,\dots,\tfrac h2-1$, an orthonormal 
basis $(\xi_{k,1},\dots,\xi_{k,(n_k)_{bc}})$ of seed states of type $k$ in $\V_{bc}$, such that for any $N$, the space $\V_{bc}(N)$ decomposes as
\begin{equation} \label{eq:decomp.Vbc.xi}
  \V_{bc}(N) = \bigoplus_{k=0,\frac 12,\dots,\frac N2}
  \bigoplus_{i=1}^{(n_k)_{bc}} \TL(N,2k)\cdot \xi_{k,i} \,.
\end{equation}
\begin{itemize}
\item For $N=0$, we have $(n_0)_{bc}=\delta_{bc}$. On the other hand,
  $$
  \V_{bc}(0) = \begin{cases}
    \mathrm{span} \, [b] & \text{if } b=c \\
    0 &\text{if } b \neq c \,.
  \end{cases}
  $$
  Moreover, if $b=c$, the state $[b]$ is automatically a seed state of type zero. Hence \eqref{eq:decomp.Vbc.xi} holds in any case.
\item For $N=1$, similarly, we have $(n_{1/2})_{bc}=N_{ab}^c$, whereas
  $$
  \V_{bc}(1) = \begin{cases}
    \mathrm{span} \, [b,c] & \text{if } N_{ab}^c=1 \\
    0 &\text{if } N_{ab}^c=0 \,.
  \end{cases}
  $$
  Moreover, if $N_{ab}^c=1$, the state $[b,c]$ is automatically a seed state of type $\tfrac 12$. Hence \eqref{eq:decomp.Vbc.xi} holds in any case.
\item For $N\geq 2$, we assume that \eqref{eq:decomp.Vbc.xi} holds for $\mathcal{V}_{bc}(N-2)$, and we consider the submodules of $\mathcal{V}_{bc}(N)$ defined as
  \begin{equation} \label{eq:V'.V''}
    \begin{aligned}
      & \mathcal{V}'_{bc}(N):= \TL(N,N-2) \cdot \mathcal{V}_{bc}(N-2) \,, \\
      & \mathcal{V}''_{bc}(N):= [\mathcal{V}'_{bc}(N)]^\perp = \{ u \in \mathcal{V}_{bc}(N) \ | \ \forall v \in \mathcal{V}'_{bc}(N) \,, \aver{u,v}=0 \} \,.
    \end{aligned}
  \end{equation}
  For any $(k,i)$ with $2k\leq N-2$, we have
  $$\TL(N,N-2)\cdot \TL(N-2,2k)\cdot\xi_{k,i} = \TL(N,2k)\cdot\xi_{k,i}\equiv I_k(N) \,,$$
  which yields
  $$
  \V'_{bc}(N) \equiv \bigoplus_{k=0,\frac 12,\dots,\frac N2-1}
  \bigoplus_{i=1}^{(n_k)_{bc}} I_k(N) \,.
  $$
  Hence, using \eqref{eq:dim.Vbc} we obtain
  $$\dim \V''_{bc}(N)=\dim \V_{bc}(N)-\dim \V'_{bc}(N)=(n_{N/2})_{bc} \,.$$
  If $N>h-2$ then $\dim \V''_{bc}(N)=0$, and thus $\V_{bc}(N)=\V_{bc}'(N)$ satisfies \eqref{eq:decomp.Vbc.xi}.
  
  Let us treat the case $N\leq h-2$.  
  For any $v \in \mathcal{V}_{bc}(N-2)$ and $j\in\{1,\dots,N-1\}$,
  the state $c_j^\dag\,v$ belongs to $\V'_{bc}(N)$.
  Thus, for any $u\in \V''_{bc}(N)$ we have $\aver{c_j\,u, v} = \aver{u, c_j^\dag\,v} = 0$. Hence, the state $c_j\,u$ is orthogonal to the module $\V_{bc}(N-2)$, and therefore $c_j\,u = 0$. This proves that any nonzero element of $\V''_{bc}(N)$ is a seed state of type $N/2$. Let $(\xi_{N/2,1},\dots,\xi_{N/2,(n_{N/2})_{bc}})$ be an orthonormal basis of $\V''_{bc}(N)$. Since the submodules $\TL(N,2k)\cdot \xi_{k,i}$ are all orthogonal to each other, they are in direct sum, which proves \eqref{eq:decomp.Vbc.xi} for $\V_{bc}(N)$.
\end{itemize}

\paragraph{Twisted periodic boundary conditions.}
The simple $\TL^a$ modules are of the form $I_{k,z}$ with $k\in\Zbb_{\geq 0}/2$ and $z\in\Cbb^\times$ (see Appendix~\ref{app:TL}).
Let us present a general method to find the decomposition of $\wh{\cal V}_{m,K}$ in the form
\begin{equation} \label{eq:decomp.VmK}
  \Vh_{m,K}(N) \equiv \bigoplus_{k=0,\frac 12,\dots,\frac{h}{2}}
  \bigoplus_{j=1}^{\nu_k(m,K)} I_{k,z_{k,j}}(N) \,,
\end{equation}
with non-negative integers $\nu_k(m,K)$, and parameters $z_{k,j} \in \Cbb^\times$ such that $|z_{k,j}|=1$.
\medskip

As a first step, using a similar argument to the one for $\dim \V_{bc}(N)$, one obtains the dimension formulas
\begin{equation} \label{eq:dim.VmK}
  \dim \Vh_{m,K}(N)
  = \begin{cases}
    \ds \alpha_0 \, D_0(N) + 2\sum_{k=\frac 12,1,\dots,\frac{h-1}{2}}  \alpha_k\, D_k(N)
    + \alpha_{\frac h2} \, D_{\frac h2}(N) & \text{if } d(a) < 2 \,, \\
    \ds \alpha_0 \, d_0(N) + 2\sum_{k=\frac 12,1,\dots,\frac{N}{2}}  \alpha_k\, d_k(N) & \text{if } d(a) \geq 2 \,,\\    
  \end{cases}
\end{equation}
where
\begin{equation}
  \alpha_k=\alpha_k(m,K)=\Tr \left[\Kh \wh{N}(m) T_{2k}\left(\frac{\wh{N}(a)}{2}\right) \right]
\end{equation}
and $T_{2k}$ is the $2k$-th Chebyshev polynomial of the first kind, whereas the dimensions $d_k(N)$ and $D_k(N)$ are defined in \eqref{eq:def.dk} and \eqref{eq:def.Dk}.

One has $\alpha_0=\Tr[\Kh\wh{N}(m)] \in \Zbb_{\geq 0}$, and one then shows by recursion that $2\alpha_k\in\Zbb$. Moreover, if $d(a)<2$, using the relation \eqref{eq:dim.VmK} at $N=h$, we get $\alpha_{h/2}\in\Zbb$.
Hence, \eqref{eq:dim.VmK} yields $\dim \wh{\cal V}_{m,K}(N)$ as a linear combination of $D_0(N),\dots,D_{h/2}(N)$ [resp. $d_0(N),\dots,d_{N/2}(N)$] with integer coefficients if $d(a)<2$ [resp. $d(a)\geq 2$].
\medskip

Next, we shall exploit the fact that any simple module over $\TL^a$ is generated by a single state:
\begin{equation}
  I_{k,z}(N) = \TL^a(N,2k) \cdot u_k \,,
\end{equation}
where $u_k$ is the unique basis state of $I_{k,z}(2k)=W_{k,z}(2k)$.
We define the seed states for $\TL^a$ modules as follows. Let $A$ be a $\TL^a$-module, $k\in\Zbb_{\geq 0}/2$ and $z\in\Cbb^\times$. A seed state of type $(k,z)$ in the module $A$ is a non-zero element $\xi\in A(2k)$ such that
\begin{equation}
  \begin{cases}
    c_1 \cdot\xi = 0 & \text{if } k\geq 1 \,, \\
    \Omega\cdot\xi=z\, \xi & \text{if } k\geq 1/2 \,, \\
    f\cdot\xi = (z+z^{-1})\,\xi & \text{if } k=0 \,.
  \end{cases}
\end{equation}
Like for the case of $\TL$ modules, if the loop weight is real and $A$ admits an invariant inner product $\aver{\,,\,}$ then the following properties hold:
\begin{itemize}
\item For any seed state $\xi$ of type $(k,z)$ with $|z|=1$, one has $\TL^a(N,2k)\cdot\xi \equiv I_{k,z}(N)$.
\item For any pair of seed states $\xi,\xi'$ of types $(k,z),(k',z')$, with $\aver{\xi,\xi'}=0$ in the case $k=k'$, the submodules $\TL^a(N,2k)\cdot\xi$ and $\TL^a(N,2k')\cdot\xi'$ are orthogonal.
\medskip
\end{itemize}

Based on these ingredients, we obtain the decomposition \eqref{eq:decomp.VmK} by a recursive argument similar to the case of $\V_{bc}(N)$. Throughout this recursion, the parameters $z_{k,j}$ are determined by the eigenvalues of $f$ and $\Omega$ respectively on $\Vh_{m,K}(0)$ and $\Vh_{m,K}(1),\Vh_{m,K}(2),\dots,\Vh_{m,K}(h)$.
\begin{itemize}
\item For $N=0$, we have $\dim \Vh_{m,K}(0)=\alpha_0=\Tr[\Kh\wh{N}(m)]$.
  Consider any orthonormal basis of eigenvectors of $f$ in $\Vh_{m,K}(0)$. By definition, if we parameterise the corresponding eigenvalues as $\varphi_j=z_{0j}+z_{0j}^{-1}$ with $z_{0j}\in\Cbb^\times$, then each of these eigenvectors is a seed state of type $(0,z_{0,j})$. We thus have
  \begin{equation} \label{eq:VmK(0)}
    \Vh_{m,K}(0) \equiv \bigoplus_{j=1}^{\nu_0(m,K)} I_{0,z_{0j}}(0) \,,
    \qquad \text{with} \quad \nu_0(m,K)=\alpha_0 \,.
  \end{equation}
  Note that, if $(m,K)=(1,\id)$, we have
  \begin{equation}
    f_a\cdot [x] = \sum_y N_{ax}^y \, [y] \,,
  \end{equation}
  and thus the eigenvalues of $f_a$ are $\varphi_j=\lambda_j(a)$.
  If $m\neq 1$ and $K=\id$, we still have the fusion relation
  \begin{equation}
    f_a \cdot f_{a'} = \sum_{a''} N_{aa'}^{a''} \, f_{a''} \,,
  \end{equation}
  and thus the eigenvalues of $f_a$ form a subset of $\{\lambda_1(a),\dots,\lambda_n(a)\}$.
  
\item For $N=1$, we have $\dim\Vh_{m,K}(1)=2\alpha_{1/2}=\Tr[\Kh\wh{N}(m)\wh{N}(a)]$. Any orthonormal basis of eigenvectors of $\Omega$ in $\Vh_{m,K}(1)$, with eigenvalues $z_{1/2,j}$, yields a basis of seed states of type $(1/2,z_{1/2,j})$.
We thus have
  \begin{equation} \label{eq:VmK(1)}
    \Vh_{m,K}(1) \equiv
    \bigoplus_{j=1}^{\nu_{1/2}(m,K)} I_{1/2,z_{j,1/2}}(1) \,,
    \qquad \text{with} \quad \nu_{1/2}(m,K)=2\alpha_{1/2} \,.
  \end{equation}
   
\item For $N\geq 2$, we assume that the decomposition \eqref{eq:decomp.VmK} holds for $\Vh_{m,K}(N-2)$, and we define the submodules of $\Vh_{m,K}(N)$:
  \begin{equation} \label{eq:VmK'}
    \Vh_{m,K}'(N) := \TL^a(N,N-2) \cdot \Vh_{m,K}(N-2) \,,
    \qquad \Vh_{m,K}''(N) := \left[\Vh_{m,K}'(N) \right]^\perp \,.
  \end{equation}
  With a similar argument to the case of $\V_{bc}$, one shows that $c_1\cdot u=0$ for any $u\in\Vh_{m,K}''(N)$.
  Consider an orthonormal basis of eigenvectors of $\Omega$ in $\Vh_{m,K}''(N)$, with eigenvalues $z_{N/2,j}$. These vectors are then seed states of type $(N/2,z_{N/2,j})$, and the submodules that they generate are orthogonal to one another.
  Hence, the decomposition \eqref{eq:decomp.VmK} holds for $\Vh_{m,K}(N)$, 
  with $\nu_{N/2}(m,K)=\dim \Vh_{m,K}''(N)$.
\end{itemize}

\section{Examples}
\label{sec:examples}

\subsection{The Fibonacci model}

\paragraph{Fusion category data.}
The Fibonacci fusion category $\rm Fib$ is defined by the simple objects $\{1,\tau\}$, with fusion rules
\begin{equation}
  1 \otimes 1 = 1 \,,
  \qquad 1 \otimes \tau = \tau \otimes 1 = \tau \,,
  \qquad \tau \otimes \tau = 1 + \tau \,.
\end{equation}
The fusion characters are $\lambda_1=(d(1),d(\tau))=(1,\phi)$ and $\lambda_2=(1,\phib)$, where $\phi=2\cos \tfrac{\pi}{5}=\tfrac 12 (1+\sqrt 5)$, and $\phib=-1/\phi = 2\cos \tfrac{3\pi}{5}$.
The $F$-symbols allowed by the fusion rules are all equal to one, except for the $F$-matrix
\begin{equation}
  F_\tau^{\tau\tau\tau} = \left(\begin{array}{cc}
    -\phib & \phi^{-1/2} \\ \phi^{-1/2} & \phib
  \end{array}\right) \,.
\end{equation}
The only category automorphism of $\rm Fib$ is the identity.
Moreover, the category is braided, with $R_{\tau\tau}=e^{-3i\pi/5}\id_{\tau\tau}+e^{3i\pi/5} \psi_1^{\tau\tau}\circ \psi^1_{\tau\tau}$.

\paragraph{The Fibonacci face model.} We consider the face model for $\C=\mathrm{Fib}$ and $a=\tau$.
The adjacency graph associated to $\Nh(\tau)$ is the tadpole diagram
\begin{equation}
  \mathfig{\includegraphics{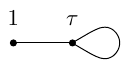}}
\end{equation}
The dominant eigenvalue is given by the quantum dimension $d(\tau)=\phi$, which yields the Coxeter number $h=5$, and the $\TL$ loop parameter $d(\tau)=-q-q^{-1}$ with $q=e^{4i\pi/5}$.
The eigenvalues of $\Nh(\tau)$ are given by $\lambda_j(\tau)=2\cos(\pi m_j/5)$, with the exponents $m_1=1$ and $m_2=3$.

If one sets $v=1$ in \eqref{eq:face} for the Boltzmann weights on every face of the lattice, one obtains a critical model, which is in the universality class of the tricritical Ising model. The face weights are given by
\begin{equation}
  \begin{array}{ccccccc}
    \sideset{_{\ds \tau}^{\ds \tau}}{_{\ds \tau}^{\ds \tau}}{\face{face.pdf}}
    & \sideset{_{\ds \tau}^{\ds \tau}}{_{\ds 1}^{\ds \tau}}{\face{face.pdf}}
    & \sideset{_{\ds \tau}^{\ds 1}}{_{\ds \tau}^{\ds \tau}}{\face{face.pdf}}
    & \sideset{_{\ds \tau}^{\ds \tau}}{_{\ds \tau}^{\ds 1}}{\face{face.pdf}}
    & \sideset{_{\ds 1}^{\ds \tau}}{_{\ds \tau}^{\ds \tau}}{\face{face.pdf}}
    & \sideset{_{\ds 1}^{\ds \tau}}{_{\ds \tau}^{\ds 1}}{\face{face.pdf}}
    & \sideset{_{\ds \tau}^{\ds 1}}{_{\ds 1}^{\ds \tau}}{\face{face.pdf}} \\
    \ \\
    2 & \phi^{-1/4} & \phi^{-1/4} & \phi^{-1/4} & \phi^{-1/4}
    & \phi^{3/2} & \phi^{3/2}
  \end{array}
\end{equation}
and they are invariant under $\pi/2$ rotations.

\paragraph{Analysis of the lattice model.}
Let us first discuss the case of fixed boundary conditions. For any $b,c$ we denote
\begin{equation}
  \V_{bc} = \V_{bc}^{\rm even} \oplus \V_{bc}^{\rm odd}
\end{equation}
as in \eqref{eq:M=Mev+Modd}.
Using the method presented in Section~\ref{sec:decomp}, we find the decompositions:
\begin{equation}
  \begin{aligned}
    \V_{11}^{\rm even} &\equiv I_0
    &\qquad& \V_{1\tau}^{\rm even} \equiv \V_{\tau 1}^{\rm even} \equiv I_1
    &\qquad& \V_{\tau\tau}^{\rm even} \equiv I_0\oplus I_1 \\
    \V_{11}^{\rm odd} &\equiv I_{\frac32}
    && \V_{1\tau}^{\rm odd} \equiv \V_{\tau 1}^{\rm odd} \equiv I_{\frac12}
    && \V_{\tau\tau}^{\rm odd} \equiv I_{\frac12}\oplus I_{\frac32}
  \end{aligned}
\end{equation}
The topological operator $X_\tau$ obeys the relation $X_\tau^2=\id+X_\tau$,
and \eqref{eq:F.I} yields its eigenvalues $\phi,\phib,\phib,\phi$ on $I_0,I_{\frac12},I_1,I_{\frac32}$ respectively. On the spaces $\V_{bc}$ with $(b,c)\neq (\tau,\tau)$, the operator $X_\tau$ is proportional to $\id$, and we have $\mathrm{End}(\V_{bc}^{\rm even})=\mathrm{span}(\id)$ and $\mathrm{End}(\V_{bc}^{\rm odd})=\mathrm{span}(\id)$. In contrast, for $(b,c)=(\tau,\tau)$, the operators $\{\id,X_\tau\}$ form a basis of both $\mathrm{End}(\V_{\tau\tau}^{\rm even})$ and $\mathrm{End}(\V_{\tau\tau}^{\rm odd})$. In all cases, the algebra generated by $\{\id,X_\tau\}$ acting on $\V_{bc}(N)$ has dimension given by
\begin{equation}
  \sum_{\smallarray{0\leq k\leq\frac h2 -1}{2k\in N+\Zbb}} [(n_k)_{bc}]^2
\end{equation}
where $(n_k)_{bc}$ is the multiplicity appearing in \eqref{eq:decomp.Vbc}.
\medskip

For periodic and twisted boundary conditions, we obtain
\begin{equation}
  \begin{aligned}
    \Vh^{\rm even} &\equiv \I_{11} \oplus \I_{33}
    &\qquad& \Vh_\tau^{\rm even} \equiv  \I_{33}  \oplus \I_{31} \oplus \I_{13} \\
    \Vh^{\rm odd} &\equiv \I_{41} \oplus \I_{23}
    &\qquad& \Vh_\tau^{\rm odd} \equiv  \I_{21} \oplus \I_{43} \oplus \I_{23}
  \end{aligned}
\end{equation}
where we have used the short-hand notation $\Vh_\tau:=\Vh_{\tau,\id}$, and the notation \eqref{eq:I} for simple modules $I_{k,(-1)^{m+k}q^m} = \I_{m+k,m-k}$.
The action of the topological operators $Y_\tau,\Yb_{\!\!\tau}$ on the simple submodules of $\Vh$ and $\Vh_\tau$
is given by \eqref{eq:F.I}. It reads
\begin{center}
  \begin{tabular}{c|cccccccc}
    & $\I_{11}$ & $\I_{33}$ & $\I_{41}$  & $\I_{23}$ & $\I_{31}$  & $\I_{13}$  & $\I_{21}$  & $\I_{43}$ \\
    \hline
    \ \\[-10pt]
    $Y_\tau$        & $\phi$ & $\phib$ & $\phi$ & $\phib$ & $\phib$ & $\phi$ & $\phib$ & $\phi$ \\
    $\Yb_{\!\!\tau}$ & $\phi$ & $\phib$ & $\phi$ & $\phib$ & $\phi$ & $\phib$ & $\phi$ & $\phib$ \\
  \end{tabular}
\end{center}
For periodic boundary conditions, the operators $\{\id,Y_\tau\}$ form a basis of both $\mathrm{End}(\Vh^{\rm even})$ and $\mathrm{End}(\Vh^{\rm odd})$, and they obey the relation $Y_\tau^2=\id+Y_\tau$, whereas $\Yb_{\!\!\tau}=Y_\tau$ on $\Vh$. For twisted boundary conditions, the operators $\{\id,Y_\tau,\Yb_{\!\!\tau}\}$ form a basis of $\mathrm{End}(\Vh_\tau^{\rm even})$ and $\mathrm{End}(\Vh_\tau^{\rm odd})$, and they obey the relations on $\Vh_\tau$
\begin{equation}
  Y_\tau^2=\id+Y_\tau \,,
  \qquad \Yb_{\!\!\tau}^2=\id+\Yb_{\!\!\tau} \,,
  \qquad Y_\tau \cdot \Yb_{\!\!\tau} = \Yb_{\!\!\tau} \cdot Y_\tau
  = -\phib^2\, \id + \phib \, (Y_\tau + \Yb_{\!\!\tau}) \,.
\end{equation}
In all cases, the algebra generated by $\{\id,Y_\tau,\Yb_{\!\!\tau}\}$ acting on $\Vh_{m,K}(N)$ has dimension
\begin{equation}
  \sum_{\smallarray{0\leq k\leq\frac h2}{2k\in N+\Zbb}} [\nu_k(m,K)]^2
\end{equation}
where $\nu_k(m,K)$ is the multiplicity appearing in \eqref{eq:decomp.VmK}.

\paragraph{Scaling limit.}
The scaling limit is described in terms of the minimal model $\mathcal{M}(5,4)$, with central charge $c=7/10$ and Kac table for the conformal dimensions $\Delta_{rs}$
\begin{center}
  \begin{tabular}{|l|l|l|l}
    \hline 
    \cellcolor{white} $\eps''$ & \cellcolor{pink}  $\sigma'$ & \cellcolor{white} $\id$    \\ \hline 
    \cellcolor{pink}  $\eps'$  & \cellcolor{white} $\sigma$  & \cellcolor{pink}  $\eps$   \\ \hline
    \cellcolor{white} $\eps$   & \cellcolor{pink}  $\sigma$  & \cellcolor{white} $\eps'$  \\ \hline
    \cellcolor{pink}  $\id$    & \cellcolor{white} $\sigma'$ & \cellcolor{pink}  $\eps''$ \\ \hline 
  \end{tabular}
\end{center}
with indices $r=1,2,3$ on the horizontal axis, and $s=1,2,3,4$ on the vertical axis.
The conformal dimensions read
\begin{equation}
  \Delta_\id=0 \,,
  \qquad \Delta_\eps=\frac{1}{10} \,,
  \qquad \Delta_{\eps'}=\frac{3}{5} \,,
  \qquad \Delta_{\eps''}=\frac{3}{2} \,,
  \qquad \Delta_\sigma=\frac{3}{80} \,,
  \qquad \Delta_{\sigma'}=\frac{7}{16} \,.
\end{equation}

The simple submodules of $\Vh^{\rm even}$ scale as
\begin{equation}
  \I_{11}\to [0,0] \oplus [\tfrac{7}{16},\tfrac{7}{16}] \oplus [\tfrac32,\tfrac32] \,,
  \qquad \I_{33} \to [\tfrac35,\tfrac35]  \oplus [\tfrac{3}{80},\tfrac{3}{80}] \oplus [\tfrac{1}{10},\tfrac{1}{10}] \,,
\end{equation}
and hence $\Vh^{\rm even}$ scales to the Hilbert space of the diagonal CFT built on $\mathcal{M}(5,4)$. For $\Vh^{\rm odd}$ we have
\begin{equation}
  \I_{41} \to [\tfrac32,0] \oplus [\tfrac{7}{16},\tfrac{7}{16}] \oplus [0,\tfrac32] \,,
  \qquad \I_{23} \to [\tfrac{1}{10},\tfrac35] \oplus [\tfrac{3}{80},\tfrac{3}{80}] \oplus [\tfrac35,\tfrac{1}{10}] \,.
\end{equation}
In this case, although the lattice model is translation invariant, in the scaling limit some of the primary operators have conformal spin $\pm \tfrac12$ or $\pm \tfrac32$. These half-integer conformal spins arise from the scaling \eqref{eq:scaling.T.Om} of the shift operator $\Omega$. Indeed, for the states of $\I_{s,\sbar}(N) \subset \Vh(N)$ which scale to $[\Delta_{rs},\Delta_{r\sbar}]$, one has $\id=\Omega^N\to (-1)^{(r+s)N}e^{-2i\pi(L_0-\Lb_0)}$. Hence, if both $N$ and $(r+s)$ are odd, the eigenvalue of $(L_0-\Lb_0)$ is a half-integer.
\medskip

For $\Vh^{\rm even}_\tau$ and $\Vh^{\rm odd}_\tau$, we have respectively
\begin{equation}
  \I_{33} \to [\tfrac35,\tfrac35]  \oplus [\tfrac{3}{80},\tfrac{3}{80}] \oplus [\tfrac{1}{10},\tfrac{1}{10}] \,,
  \quad \I_{31} \to [\tfrac35,0]  \oplus [\tfrac{3}{80},\tfrac{7}{16}] \oplus [\tfrac{1}{10},\tfrac32] \,,
  \quad \I_{13} \to [0,\tfrac35]  \oplus [\tfrac{7}{16},\tfrac{3}{80}] \oplus [\tfrac32,\tfrac{1}{10}] \,,
\end{equation}
and
\begin{equation}
  \I_{21} \to [\tfrac{1}{10},0] \oplus [\tfrac{3}{80},\tfrac{7}{16}] \oplus [\tfrac35,\tfrac32] \,,
  \quad \I_{43} \to [\tfrac32,\tfrac35]  \oplus [\tfrac{7}{16},\tfrac{3}{80}] \oplus [0,\tfrac{1}{10}] \,,
  \quad \I_{23} \to [\tfrac{1}{10},\tfrac35]  \oplus [\tfrac{3}{80},\tfrac{3}{80}] \oplus [\tfrac35,\tfrac{1}{10}] \,.
\end{equation}
Here, the twisted boundary conditions allow fractional conformal spins.

\subsection{The Ising model}

\paragraph{Fusion category data.}
The Ising fusion category is defined by three objects $\{+,\nu,-\}$, with fusion rules given by
\begin{equation}
  \begin{aligned}
    & [+] \otimes [+] = [+] \,,
    &\quad& [+] \otimes [-] = [-] \otimes [+] = [-] \,,
    \quad [-] \otimes [-] = [+] \,, \\
    & [\nu] \otimes [\nu] = [+] + [-] \,,
    &\quad& [\nu] \otimes [\pm] = [\pm] \otimes [\nu] = [\nu] \,.
  \end{aligned}
\end{equation}
In particular, $1=[+]$ is the identity object, whereas $[-]$ acts as the $\Zbb_2$ spin flip.
The fusion characters are
\begin{equation} \label{eq:fusion.char.ising}
  \lambda_1=(1,\sqrt 2,1), \qquad \lambda_2=(1,0,-1), \qquad \lambda_3=(1,-\sqrt 2,1) \,.
\end{equation}
The $F$-symbols can be found in \cite{KL94,Guide}.
The only category automorphism is the identity.
The category is braided, and we have $R_{\nu\nu}=q^{1/2}\id_{\nu\nu}+q^{-1/2} \psi_1^{\nu\nu}\circ \psi^1_{\nu\nu}$.

\paragraph{The Ising model as a face model.} We consider the face model with $a=\nu$. The adjacency graph associated to $G=\wh{N}(\nu)$ is the Dynkin diagram
\begin{equation} \label{eq:A3}
  A_3 = \quad \mathfig{\includegraphics{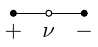}}
\end{equation}
with dominant eigenvalue $\sqrt 2$, which yields the Coxeter number $h=4$. The corresponding TL loop weight is $\beta=-q-q^{-1}$, with $q=e^{3i\pi/4}$.
The eigenvalues of $\Nh(\nu)$ read $\lambda_j(\nu)=2\cos(\pi j/4)$, with $j=1,2,3$.

On the diagram \eqref{eq:A3}, we have indicated the bipartition in to black ($\pm$) and white ($\nu$) objects. Due to this bipartition, in any spin configuration, half of the spins are ``frozen'' to the value $\nu$, and the other spins take values $\pm$.
Upon rescaling by an overall factor $2^{1/4}$, the face weights \eqref{eq:face} read
\begin{equation}
  \begin{array}{ccccccc}
    \sideset{_{\ds \pm}^{\ds \nu}}{_{\ds \nu}^{\ds \pm}}{\face{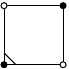}}
    & \sideset{_{\ds \pm}^{\ds \nu}}{_{\ds \nu}^{\ds \mp}}{\face{face.bw.pdf}} 
    & \sideset{_{\ds \nu}^{\ds \pm}}{_{\ds \pm}^{\ds \nu}}{\face{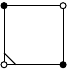}} 
    & \sideset{_{\ds \nu}^{\ds \pm}}{_{\ds \mp}^{\ds \nu}}{\face{face.wb.pdf}} \\[20pt]
    \sqrt2 + v & v & 1+v \sqrt2 & 1
  \end{array}
\end{equation}
We set $v=v_1$ and $v=v_2$ respectively on the faces of type $\face{face.bw.pdf}$ and $\face{face.wb.pdf}$.
The face model is then equivalent to an Ising model with spins $x_i=\pm 1$ on the black sublattice, with Boltzmann weights
$\exp(J_1 x_ix_j)$ and $\exp(J_2 x_ix_j)$ respectively on the two types of edges,
where the coupling constants read
\begin{equation}
  \tanh J_1 = \frac{1}{1+v_1 \sqrt2} \,,
  \qquad \tanh J_2 = \frac{1}{1+ v_2^{-1}\sqrt2} \,.
\end{equation}
The Kramers--Wannier self-dual line is given by $v_1=v_2$, or equivalently $\sinh 2J_1 \sinh 2J_2=1$. The model is critical along this line.
In the anisotropic limit $J_2\to 0$ on the self dual line, the transfer matrix reduces to the critical quantum Ising Hamiltonian
\begin{equation} \label{eq:H.Ising} 
  H = -\sum_{j=1}^{N/2}(\sigma_j^z\sigma_{j+1}^z + \sigma_j^x) \,,
\end{equation}
where $\sigma_j^\alpha$ denotes the Pauli matrix $\sigma^\alpha$ acting on the spin $j$ of the quantum chain.

\paragraph{Analysis of the lattice model.}
For open boundary conditions, the spin values $\pm$ for the face model correspond to fixed boundary conditions $\pm$ for the Ising model, whereas the spin value $\nu$ yields free ($f$) boundary conditions.
Using the method presented in Section~\ref{sec:decomp}, we obtain the decompositions:
\begin{equation}
  \begin{aligned}
    & \V_{++} \equiv \V_{--} \equiv I_0
    &\qquad& \V_{ff}\equiv I_0 \oplus I_1  \\
    & \V_{\pm f} \equiv \V_{f\pm} \equiv I_{\frac12}
    && \V_{+-} \equiv \V_{-+} \equiv I_1 \,.
  \end{aligned}
\end{equation}
Due to the bipartition of the graph \eqref{eq:A3}, one has
\begin{equation}
  \begin{aligned}
    & \V_{++}(N) = \mathrm{span} \left(
      [+,\nu,x_2,\nu,x_4,\dots,\nu,+] \,,
      \quad x_2,x_4,\dots,x_{N-2} = \pm
    \right) \,, \\
    & \V_{f+}(N) = \mathrm{span} \left(
      [\nu,x_1,\nu,x_3,\dots,\nu,+] \,,
      \quad x_1,x_3,\dots,x_{N-2} = \pm
    \right) \,,
  \end{aligned}
\end{equation}
and similarly for the other spaces $\V_{bc}$.
The operator $Y_-$ flips the Ising spins:
\begin{equation}
  Y_- \cdot [x_0,\nu,x_2,\nu,\dots] = [-x_0,\nu,-x_2,\nu,\dots] \,,
  \qquad Y_- \cdot [\nu,x_1,\nu,x_3,\dots] = [\nu,-x_1,\nu,-x_3,\dots] \,.
\end{equation}
The operator $Y_\nu$ inserts a seam with elementary weights
\begin{equation} \label{eq:seam.Ising}
  \begin{aligned}
  \begin{array}{cccc}
    \sideset{_{\ds +}^{\ds \nu}}{_{\ds \nu}^{\ds +}}{\face{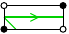}}
    & \sideset{_{\ds +}^{\ds \nu}}{_{\ds \nu}^{\ds -}}{\face{seam.bw.pdf}}
    & \sideset{_{\ds -}^{\ds \nu}}{_{\ds \nu}^{\ds +}}{\face{seam.bw.pdf}}
    & \sideset{_{\ds -}^{\ds \nu}}{_{\ds \nu}^{\ds - }}{\face{seam.bw.pdf}} \\[8pt]
    \frac{1}{\sqrt 2} & \frac{1}{\sqrt 2} & \frac{1}{\sqrt 2} & -\frac{1}{\sqrt 2}
  \end{array} \\[15pt]
  \begin{array}{cccc}
    \sideset{^{\ds +}_{\ds \nu}}{^{\ds \nu}_{\ds +}}{\face{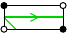}}
    & \sideset{^{\ds +}_{\ds \nu}}{^{\ds \nu}_{\ds -}}{\face{seam.wb.pdf}}
    & \sideset{^{\ds -}_{\ds \nu}}{^{\ds \nu}_{\ds +}}{\face{seam.wb.pdf}}
    & \sideset{^{\ds -}_{\ds \nu}}{^{\ds \nu}_{\ds - }}{\face{seam.wb.pdf}} \\[8pt]
    1 & 1 & 1 & -1
  \end{array}
  \end{aligned}
\end{equation}
Since the Ising category is braided, we shall describe the algebra of topological operators in terms of $\id,X_\nu,X_-$ instead of $\id,Y_\nu,Y_-$. These operators satisfy the fusion relations
\begin{equation}
  X_-^2 =\id \,, \qquad X_\nu^2=\id+X_- \,,
  \qquad X_-X_\nu = X_\nu X_- = X_\nu \,.
\end{equation}
From \eqref{eq:F.I}, the operator $X_\nu$ has eigenvalues $\sqrt 2,0,-\sqrt 2$ respectively on $I_0,I_{1/2},I_1$, and we have $X_-=X_\nu^2-\id$. For any $(b,c)\neq(f,f)$, one has $\mathrm{End}(\V_{bc})=\mathrm{span}(\id)$. On $\V_{ff}$ one has $X_-=\id$, and the operators $(\id,X_\nu)$, subject to $X_\nu^2/2=\id$, form a basis of $\mathrm{End}(\V_{ff})$.
\medskip

Let us turn to closed boundary conditions. The twist labels $+$ and $-$ yield respectively periodic and anti-periodic boundary conditions on the Ising spin, whereas the twist label $\nu$ amounts to inserting a path with weights \eqref{eq:seam.Ising}.
The decomposition of the space of states reads
\begin{equation}
  \Vh \equiv \I_{11} \oplus \I_{22} \oplus \I_{33} \,, 
  \qquad \Vh_- \equiv \I_{22} \oplus \I_{31} \oplus \I_{13} \,, 
  \qquad \Vh_\nu \equiv \I_{21} \oplus \I_{12} \oplus \I_{32} \oplus \I_{23} \,,
\end{equation}
with the notation $\Vh_m:=\Vh_{m,\id}$, and $I_{k,(-1)^{m+k}q^m}=\I_{m+k,m-k}$ as in \eqref{eq:I}.
The topological operators $\id,Y_\nu,Y_-$ satisfy the fusion relations
\begin{equation}
  Y_-^2 =\id \,, \qquad Y_\nu^2=\id+Y_- \,,
  \qquad Y_-Y_\nu = Y_\nu Y_- = Y_\nu \,,
\end{equation}
and similarly for $\id,\Yb_{\!\!\nu},\Yb_{\!\!-}$. 
The operators $Y_\nu$ and $\Yb_{\!\!\nu}$ have eigenvalues on the simple submodules of $\Vh_m$ given by \eqref{eq:F.I}, which read
\begin{center}
  \begin{tabular}{c|ccccccccc}
    & $\I_{11}$ & $\I_{22}$ & $\I_{33}$ & $\I_{31}$ & $\I_{13}$ & $\I_{21}$ & $\I_{12}$ & $\I_{32}$ & $\I_{23}$   \\
    \hline
    \ \\[-10pt]
    $Y_\nu$        & $\sqrt 2$ & $0$ & $-\sqrt 2$ & $-\sqrt 2$ & $\sqrt 2$ & $0$ & $\sqrt 2$ & $-\sqrt 2$ & $0$ \\
    $\Yb_{\!\!\nu}$ & $\sqrt 2$ & $0$ & $-\sqrt 2$ & $\sqrt 2$  & $-\sqrt 2$ & $-\sqrt 2$ & $0$ & $0$ & $\sqrt 2$ \\
  \end{tabular}
\end{center}
The algebras of topological operators on the spaces $\Vh_m$ are described in the following table.
\begin{center}
  \begin{tabular}{c|c|c}
    module $A$ & basis of $\mathrm{End}(A)$ & additional relations on $A$  \\
    \hline
    \ && \\[-10pt]
    $\Vh_{\ }$ & $\{\id,Y_\nu,Y_-\}$ & $\Yb_{\!\!\nu}=Y_\nu$ and $\Yb_{\!\!-}=Y_-$ \\
    $\Vh_-$ & $\{\id,Y_\nu,Y_-\}$ & $\Yb_{\!\!\nu}=-Y_\nu$ and $\Yb_{\!\!-}=Y_-$ \\
    $\Vh_\nu$ & $\{\id,Y_\nu,\Yb_{\!\!\nu},Y_-\}$ & $\Yb_{\!\!\nu}Y_\nu=Y_\nu\Yb_{\!\!\nu}=0$ and $\Yb_{\!\!-}=-Y_-$
  \end{tabular}
\end{center}

\paragraph{Scaling limit.}
The Ising model on the self-dual line scales to the minimal CFT $\mathcal{M}(4,3)$ with
central charge $c=1/2$ and Kac table for the conformal dimensions $\Delta_{rs}$
\begin{center}
  \begin{tabular}{|l|l|l|}
    \hline 
    \cellcolor{pink}  $\eps$   & \cellcolor{white} $\id$    \\ \hline
    \cellcolor{white} $\sigma$ & \cellcolor{pink}  $\sigma$ \\ \hline
    \cellcolor{pink}  $\id$    & \cellcolor{white} $\eps$   \\ \hline 
  \end{tabular}
\end{center}
with indices $r=1,2$ on the horizontal axis, and $s=1,2,3$ on the vertical axis.
The conformal dimensions read
\begin{equation}
  \Delta_\id = 0 \,,
  \qquad \Delta_\sigma=\frac{1}{16} \,,
  \qquad \Delta_\eps=\frac{1}{2} \,.
\end{equation}
Using the scaling limit \eqref{eq:scaling.I} of the simple modules, we obtain
\begin{equation} \label{eq:V.Ising}
  \begin{aligned}
    &\Vh \  \to 2([0,0] \oplus [\tfrac1{16},\tfrac1{16}] \oplus [\tfrac12,\tfrac12]) \,, \\
    &\Vh_- \to 2([\tfrac12,0] \oplus [0,\tfrac12] \oplus [\tfrac1{16},\tfrac1{16}]) \,, \\
    &\Vh_\nu \to 2([\tfrac1{16},0] \oplus [0,\tfrac1{16}] \oplus [\tfrac1{16},\tfrac12] \oplus [\tfrac12,\tfrac1{16}]) \,.
  \end{aligned}
\end{equation}

Let us first discuss the case of periodic boundary conditions.
Due to the bipartition of the graph \eqref{eq:A3}, $\Vh(N)$ decomposes as a vector space (but not as a $\TL^a$ module) in two subspaces
\begin{equation} \label{eq:Vh.bw}
  \begin{aligned}
    &\Vh^\bullet(N) =  \mathrm{span}\left([x_0,x_1,\dots,x_N] \,| x_0=x_N\neq \nu \right) \,, \\
    & \Vh^\circ(N) =  \mathrm{span}\left([x_0,x_1,\dots,x_N] \,| x_0=x_N=\nu \right) \,.
    \end{aligned}
\end{equation}
Each subspace carries a single copy of the Ising model.
Denote by $\ket\id$ the ground state of $H$~\eqref{eq:H.Ising}, say in $\Vh^\bullet(N)$. From the scaling \eqref{eq:scaling.T.Om} of $\Omega$, we can identify $(\id+\Omega)\ket\id$ and $(\id-\Omega)\ket\id$ as the ground states in $\I_{11}(N)$ and $\I_{33}(N)$, respectively. Hence, the topological operators act on these states as
\begin{equation}
  Y_- \cdot (\id\pm\Omega)\ket\id = (\id\pm\Omega)\ket\id \,,
  \qquad Y_\nu \cdot (\id\pm\Omega)\ket\id = \pm\sqrt 2 \,(\id\pm\Omega)\ket\id \,,
\end{equation}
which yields $Y_- \cdot \ket{\id} = \ket{\id}$ and $Y_X \cdot \ket{\id} = \sqrt 2 \, \Omega\ket{\id}$.
Similarly, let $\ket\sigma$ (resp. $\ket\eps$) be the eigenstate of $H$ \eqref{eq:H.Ising} in $\Vh^\bullet(N)$  which scales to the primary state $\ket{\tfrac1{16},\tfrac1{16}}$ (resp. $\ket{\tfrac12,\tfrac12}$) as $N \to \infty$.
With the same argument as for $\ket{\id}$, we obtain
\begin{equation}
  \begin{aligned}
    & Y_- \cdot \ket\id = +\ket\id
    &\qquad& Y_\nu \cdot \ket\id = + \sqrt 2 \, \Omega\ket\id \\
    & Y_- \cdot \ket\sigma = -\ket\sigma
    && Y_\nu \cdot \ket\sigma = 0 \\
    & Y_- \cdot \ket\eps = +\ket\eps
    && Y_\nu \cdot \ket\eps = -\sqrt 2 \, \Omega\ket\eps
  \end{aligned}
\end{equation}
The above action of $Y_-$ on $\ket\id,\ket\sigma,\ket\eps$ is consistent with the fact that $Y_-$ is, by construction, the $\Zbb_2$ spin flip operator. The operator $Y_\nu$ exchanges the roles of the two sublattices for the face model, and it yields a minus sign when applied to $\ket\eps$. Hence, inserting $Y_\nu$ along a contour can be interpreted as applying the Kramers--Wannier duality to the interior of this contour.
\medskip

For antiperiodic boundary conditions, the three sectors of $\Vh_-$ in \eqref{eq:V.Ising} are generated  respectively by the primary states corresponding to the chiral fermions $\psi,\psib$ and the disorder operator $\mu$.
Using the same arguments as for periodic boundary conditions, we obtain
\begin{equation}
  \begin{aligned}
    & Y_-\cdot\ket\psi= +\ket\psi
    &\qquad& Y_\nu \cdot\ket\psi= -\sqrt2\,\Omega\ket\psi \\
    & Y_-\cdot\ket\psib= +\ket\psib
    && Y_\nu \cdot\ket\psib=  +\sqrt2\,\Omega\ket\psib \\
    & Y_-\cdot\ket\mu= -\ket\mu
    && Y_\nu \cdot\ket\mu= 0 
  \end{aligned}
\end{equation}
whereas $\Yb_{\!\!-}=Y_-$ and $\Yb_{\!\!\nu}=-Y_\nu$ on $\Vh_-$.
\medskip

Finally, the four sectors of $\Vh_\nu$ in \eqref{eq:V.Ising} correspond to the primary operators which we denote as $\chi,\chib,\eta,\etab$. The topological operators act as
\begin{equation}
  \begin{aligned}
    & Y_-\cdot\ket\chi= -\ket\chi
    &\qquad& Y_\nu\cdot\ket\chi= 0
    &\qquad& \Yb_{\!\!\nu}\cdot\ket\chi= +\sqrt2 \, \Omega\ket\chi \\
    & Y_-\cdot\ket\chib= +\ket\chib
    && Y_\nu\cdot\ket\chib= +\sqrt2 \, \Omega\ket\chib
    && \Yb_{\!\!\nu}\cdot\ket\chib=0 \\
    & Y_-\cdot\ket\eta= -\ket\eta
    && Y_\nu\cdot\ket\eta=0
    && \Yb_{\!\!\nu}\cdot\ket\eta=  -\sqrt2 \, \Omega\ket\eta \\
    & Y_-\cdot\ket\etab= +\ket{\etab}
    && Y_\nu\cdot\ket\etab=-\sqrt2 \, \Omega\ket\etab
    && \Yb_{\!\!\nu}\cdot\ket\etab=0
  \end{aligned}
\end{equation}
whereas $\Yb_{\!\!-}=-Y_-$ on $\Vh_\nu$.

\subsection{The three-state Potts model}

\paragraph{The TY($\Zbb_3$) category.}
The Tambara--Yamagami fusion category associated to the $\Zbb_3$ group is defined as follows \cite{TY98}.
The simple objects are $\{1,\omega,\omegab,\nu\}$, and the fusion rules read
\begin{equation}
  \begin{tabular}{c|cccc}
    $\otimes$ & $1$ & $\omega$ & $\omegab$ & $\nu$ \\
    \hline
    $1$ & $1$ & $\omega$ & $\omegab$ & $\nu$ \\
    $\omega$ & $\omega$ & $\omegab$ & $1$ & $\nu$ \\
    $\omegab$ & $\omegab$ & $1$ & $\omega$ & $\nu$\\
    $\nu$ & $\nu$ & $\nu$ & $\nu$ & $1+\omega+\omegab$
  \end{tabular}
\end{equation}
The simple object $\omega$ acts as the $\Zbb_3$ permutation $(1,\omega,\omegab,\nu)\mapsto (\omega,\omegab,1,\nu)$, and $\omegab$ is its inverse.
The fusion characters read
\begin{equation}
  \lambda_1=(1,1,1,\sqrt 3) \,,
  \quad \lambda_2=(1,\omega,\omegab,0) \,,
  \quad \lambda_3=(1,\omegab,\omega,0) \,,
  \quad \lambda_4=(1,1,1,-\sqrt 3) \,,
\end{equation}
where $\omega=e^{2i\pi/3}$ and $\omegab=e^{-2i\pi/3}$.
The $F$-symbols can be found in \cite{TY98}.
The $\Zbb_2$ permutation $\zeta:p \mapsto \pb$ is a category automorphism.
The category TY($\Zbb_3$) is not braided.

\paragraph{The three-state Potts model.}
The relation between the three-state Potts model and the face model with $\C=$TY($\Zbb_3$) and $a=\nu$ is similar to the case of the Ising model. The adjacency matrix $G=\wh{N}(\nu)$ corresponds to the Dynkin diagram
\begin{equation} \label{eq:D4}
  D_4 = \quad \mathfig{\includegraphics{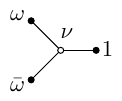}}
\end{equation}
with Coexeter number $h=6$, which yields $q=e^{5i\pi/6}$.
The eigenvalues of $\Nh(\nu)$ read $\lambda_j(\nu)=2\cos(\pi m_j/6)$ with exponents $m_1,\dots,m_4=1,3,3,5$.

Using the same conventions as for the Ising model, one has the edge interactions $\exp(J_1\,\delta_{x_i,x_j})$ and $\exp(J_2\,\delta_{x_i,x_j})$ on the two types of edges respectively, where $J_1=\log(1+\sqrt{3}/x_1)$ and $J_2=\log(1+x_2\sqrt{3})$. The self-dual line is $v_1=v_2$, corresponding to $(e^{J_1}-1)(e^{J_2}-1)=3$. The isotropic self-dual point is $J_1=J_2=\log(1+\sqrt{3})$.

\paragraph{Analysis of the lattice model.}
For open boundary conditions, we consider the fixed boundary conditions $1,\omega,\omegab$ and free boundary conditions $f$ corresponding to the simple object $\nu$.
We obtain the decompositions
\begin{equation}
  \begin{aligned}
    & \V_{11} \equiv \V_{\omega\omega}\equiv \V_{\omegab\omegab}\equiv  I_0 \oplus I_2 \,,
    &\qquad& \V_{1\omega}\equiv \V_{1\omegab}\equiv \V_{\omega\omegab}\equiv  I_1 \,, \\
    & \V_{1f}\equiv \V_{\omega f}\equiv \V_{\omegab f}\equiv  I_{\frac12} \oplus I_{\frac32} \,,
    && \V_{ff} \equiv  I_0 \oplus 2 I_1 \oplus I_2 \,,
  \end{aligned}
\end{equation}
and the other decompositions are obtained by the relation $\V_{bc} \equiv \V_{cb}$.
The topological operators $Y_1,Y_\omega,Y_{\omegab},Q_\zeta$ act on the basis states as
\begin{equation}
  \begin{aligned}
    & Y_p \cdot [x_0,\nu,x_2,\nu,\dots]
    = [p\cdot x_0,\nu,p\cdot x_2,\nu,\dots] \,, \qquad p=1,\omega,\omegab \\
    & Q_\zeta \cdot [x_0,\nu,x_2,\nu,\dots] = [\xb_0,\nu,\xb_2,\nu,\dots] \,,
  \end{aligned}
\end{equation}
and similarly for the states $[\nu,x_1,\nu,x_3,\dots]$.
The operator $Y_\nu$ inserts a seam with elementary weights
\begin{equation} \label{eq:seam.Potts}
  \begin{array}{ccc}
    \sideset{_{\ds \omega^i}^{\ds \nu}}{_{\ds \nu}^{\ds \omega^j}}{\face{seam.bw.pdf}}
    &\qquad\qquad& \sideset{_{\ds \nu}^{\ds \omega^i}}{_{\ds \omega^j}^{\ds \nu}}{\face{seam.wb.pdf}}\\[10pt]
    \ds\frac{\omega^{-ij}}{\sqrt 3} && \omega^{ij}
  \end{array}
\end{equation}
Note that $Y_\nu$ vanishes on all the modules $\V_{bc}$, whereas $Y_\omega$ and $Q_\zeta$ generate the dihedral group $\mathbb{D}_3$.
Since TY($\Zbb_3$) is not braided, we cannot rely on \eqref{eq:F.I} to compute the action of the topological operators on the simple submodules. Instead, use the seed states $\xi_k$, on which the action of the operators is easy to determine.
The results are shown in the following table.
\begin{center}
  \begin{tabular}{c|c|c}
    module $A$ & seed states & basis of $\mathrm{End}(A)$ \\
    \hline
    \ && \\[-10pt]
    $\V_{1\omega}$ & $\xi_1=[1,\nu,\omega]$ & $\id$  \\
    \hline
    $\V_{11}$ & $\xi_0=[1],\ \xi_2=[1,\nu,\omega,\nu,1]-[1,\nu,\omegab,\nu,1]$ & $\id,Q_\zeta$ \\
    \hline
    $\V_{1f}$ & $\xi_{\frac12}=[1,\nu],\ \xi_{\frac32}=[1,\nu,\omega,\nu]-[1,\nu,\omegab,\nu]$ & $\id,Q_\zeta$ \\
    \hline
    $\V_{ff}$ & $\begin{array}{c}
      \xi_0=[\nu] \\
      \xi'_1=[\nu,1,\nu]+\omega[\nu,\omega,\nu]+\omegab[\nu,\omegab,\nu] \\
      \xi''_1=[\nu,1,\nu]+\omegab[\nu,\omega,\nu]+\omega[\nu,\omegab,\nu] \\
      \xi_2=[\nu,1,\nu,\omega,\nu]-[\nu,1,\nu,\omegab,\nu]+[\nu,\omega,\nu,\omegab,\nu] \\
      \qquad -[\nu,\omega,\nu,1,\nu]+[\nu,\omegab,\nu,1,\nu]-[\nu,\omegab,\nu,\omega,\nu]
    \end{array}$ & $\id,Y_\omega,Y_{\omegab},Q_\zeta,Q_\zeta Y_\omega,Q_\zeta Y_{\omegab}$
  \end{tabular}
\end{center}
\medskip

For periodic and twisted boundary conditions, we obtain the decompositions
\begin{equation} \label{eq:decomp.Potts}
  \begin{aligned}
    & \Vh \equiv
    \I_{11} \oplus 2\,\I_{33} \oplus \I_{55} \oplus \I_{51} \oplus \I_{15} \,, \\
    & \Vh_{\omega,\id} \equiv \Vh_{\omegab,\id}
    \equiv \I_{33} \oplus \I_{31} \oplus \I_{13} \oplus \I_{53} \oplus \I_{35} \,, \\
    & \Vh_{\nu,\id} \equiv \I_{21} \oplus \I_{41}
    \oplus 2\, \I_{23} \oplus 2\, \I_{43} \oplus \I_{25} \oplus \I_{45} \,, \\
    & \Vh_{1,\zeta}\equiv \Vh_{\omega,\zeta}\equiv \Vh_{\omegab,\zeta}
    \equiv \I_{22} \oplus \I_{44} \oplus \I_{24} \oplus \I_{42} \,, \\
    & \Vh_{\nu,\zeta}\equiv \I_{12} \oplus \I_{14}
    \oplus 2\, \I_{32} \oplus 2\, \I_{34} \oplus \I_{52} \oplus \I_{54} \,.
  \end{aligned}
\end{equation}
Let us discuss the topological operators on the space $\Vh$ with periodic boundary conditions.
Recall that $\I_{mm}=I_{0,(-q)^m}=I_{0,\exp(i\pi m/h)}$, and thus each character $\lambda_j$ yields a seed state for the submodule $\I_{m_j,m_j}$, where $m_j$ is the exponent associated to $\lambda_j$.
Since for any $p$, the operator $Y_p$ acts as $\Nh(p)$ on $\Vh(0)$, we have $Y_p \cdot \lambda_j = \lambda_j(p)\,\lambda_j$, and thus $Y_p$ acts as $\lambda_j(p)\id$ on the submodule $\I_{m_j,m_j}$. For the submodules $\I_{51}$ and $\I_{15}$ of $\Vh$, the seed states read respectively
\begin{equation}
  \begin{aligned}
    & \xi_{51} = (\id+i\Omega-\Omega^2-i\Omega^3) \,
    (\id+Y_\omega+Y_{\omegab}) \cdot [\nu,1,\nu,\omega,\nu] \,, \\
    & \xi_{15} = (\id-i\Omega-\Omega^2+i\Omega^3) \,
    (\id+Y_\omega+Y_{\omegab}) \cdot [\nu,1,\nu,\omega,\nu] \,.
  \end{aligned}
\end{equation}
We thus get the action of the topological operators on the submodules of $\Vh$:
\begin{center}
  \begin{tabular}{c|ccccc}
    & $\I_{11}$ & $2\, \I_{33}$ & $\I_{55}$ & $\I_{51}$ & $\I_{15}$ \\
    \hline
    $Y_\omega$ & $1$ & $\left(\begin{array}{cc} \omega & 0 \\ 0 & \omegab \end{array}\right)$ & $1$ & $1$ & $1$ \\
    $Y_{\omegab}$ & $1$ &$\left(\begin{array}{cc} \omegab & 0 \\ 0 & \omega \end{array}\right)$& $1$ & $1$ & $1$ \\
    $Y_\nu$ & $\sqrt3$ & $\left(\begin{array}{cc} 0 & 0 \\ 0 & 0 \end{array}\right)$& $-\sqrt3$ & $\sqrt3$ & $-\sqrt3$ \\
    $Q_\zeta$ & 1 & $\left(\begin{array}{cc} 0 & 1 \\ 1 & 0 \end{array}\right)$& $1$ & $-1$ & $-1$
  \end{tabular}
\end{center}
One can check explicitly that the eight topological operators $Y_p$ and $Q_\zeta Y_p$ with $p=1,\omega,\omegab,\nu$ are linearly independent, and hence they form a basis of $\mathrm{End}(\Vh)$.
\medskip

A similar analysis may be carried out for the spaces with twisted boundary conditions, up to replacing the operators $Y_p$ by the appropriate $Y_{p,\alpha}$ as in \eqref{eq:Ypa} -- we leave this as an exercise for the interested readers.

\paragraph{Scaling limit.}
The three-state Potts model on the self-dual line scales to the minimal CFT $\mathcal{M}(6,5)$ with central charge $c=4/5$ and Kac table for conformal dimensions $\Delta_{rs}$
\begin{center}
  \begin{tabular}{|l|l|l|l|l}
    \hline
    \cellcolor{pink}  $Y$   & \cellcolor{white} $X$ &
    \cellcolor{pink}  $\eps$   & \cellcolor{white} $\id$    \\ \hline
    \cellcolor{white}  $\phi_{42}$  & \cellcolor{pink} $\phi_{32}$
    & \cellcolor{white} $\phi_{22}$  & \cellcolor{pink}  $\phi_{12}$  \\ \hline 
    \cellcolor{pink}  $Z$ & \cellcolor{white} $\sigma$
    & \cellcolor{pink}  $\sigma$ & \cellcolor{white} $Z$ \\ \hline
    \cellcolor{white}  $\phi_{12}$  & \cellcolor{pink}  $\phi_{22}$ 
    & \cellcolor{white} $\phi_{32}$  & \cellcolor{pink}  $\phi_{42}$  \\ \hline 
    \cellcolor{pink}  $\id$    & \cellcolor{white} $\eps$
    & \cellcolor{pink} $X$  & \cellcolor{white} $Y$   \\ \hline 
  \end{tabular}
\end{center}
with indices $r=1,2,3,4$ on the horizontal axis, and $s=1,2,3,4,5$ on the vertical axis.
The conformal dimensions read
\begin{equation}
  \Delta_\id = 0 \,,
  \qquad \Delta_\eps=\frac25 \,,
  \qquad \Delta_\sigma=\frac{1}{15} \,,
  \qquad \Delta_X=\frac75 \,,
  \qquad \Delta_Y=3 \,,
  \qquad \Delta_Z= \frac23 \,,
\end{equation}
and
\begin{equation}
  \Delta_{12}= \frac18 \,,
  \qquad \Delta_{22}= \frac1{40} \,,
  \qquad \Delta_{32}= \frac{21}{40} \,,
  \qquad \Delta_{42}= \frac{13}8 \,.
\end{equation}
Using the scaling
$$\I_{s\sbar}\to \bigoplus_{r=1}^4 [\Delta_{rs},\Delta_{r\sbar}]$$
we see that the decompositions \eqref{eq:decomp.Potts} correspond to the block structure found in \cite{PZ01}. Due to the bipartition of the graph $D_4$, the spaces of states $\Vh_{m,K}$ decompose in two subspaces $\Vh_{m,K}^\bullet$ and $\Vh_{m,K}^\circ$ as in \eqref{eq:Vh.bw}.

For periodic boundary conditions, the scalar primary states are
$$\ket\id,\ \ket\eps,\ \ket\sigma,\ \ket{\sigma^\dag},\ \ket{X},\ \ket{Y},\ \ket{Z},\ \ket{Z^\dag}$$
and the non-scalar states are $\ket{W},\ket{\overline  W},\ket{\eps'},\ket{\bar \eps'}$, associated respectively to the modules $[3,0]$, $[0,3]$, $[\tfrac75,\tfrac25]$ and $[\tfrac25,\tfrac75]$.
We obtain the action of the topological operators through an analysis similar to the Ising case.
We get:
\begin{equation}
  Y_\omega\cdot \ket\sigma = \omega\, \ket\sigma \,,
  \quad Y_\omega\cdot \ket{\sigma^\dag} =  \omegab\, \ket\sigma \,,
  \quad Y_\omega\cdot \ket{Z} = \omega\, \ket{Z} \,,
  \quad Y_\omega\cdot \ket{Z^\dag} = \omegab\,  \ket{Z^\dag} \,,
\end{equation}
and the other primary states are invariant under $Y_\omega$.
For $Y_\nu$, we get:
\begin{equation}
  \begin{aligned}
    &Y_\nu\cdot \ket\id = \sqrt 3\, \Omega\ket\id \,,
    &\qquad& Y_\nu\cdot \ket\eps = -\sqrt 3\, \Omega\ket\eps \,,
    &\qquad& Y_\nu\cdot \ket\sigma = Y_\nu\cdot \ket{\sigma^\dag} = 0 \,, \\
    & Y_\nu\cdot \ket{X} = \sqrt 3\, \Omega\ket{X} \,,
    && Y_\nu\cdot \ket{Y} = -\sqrt 3\, \Omega\ket{Y} \,,
    && Y_\nu\cdot \ket{Z} = Y_\nu\cdot \ket{Z^\dag} = 0 \,,
  \end{aligned}
\end{equation}
and
\begin{equation}
  \begin{aligned}
    & Y_\nu\cdot \ket{W} = \sqrt 3\, \Omega\ket{W} \,,
    &\qquad& Y_\nu\cdot \ket{\overline W} = -\sqrt 3\, \Omega\ket{\overline W} \,, \\
    & Y_\nu\cdot \ket{\eps'} = -\sqrt 3\, \Omega\ket{\eps'} \,,
    && Y_\nu\cdot \ket{\bar\eps'} = \sqrt 3\, \Omega\ket{\bar\eps'} \,.
  \end{aligned}
\end{equation}

\subsection{A psu(2)$_5$ model}

\paragraph{Fusion category data.}
The psu$(2)_5$ category has three simple objects $\{1,2,3\}$, and fusion rules
\begin{equation}
  [2] \times [2] = [1]+[3] \,,
  \qquad [2] \times [3] = [3] \times [2] = [2]+[3] \,,
  \qquad [3] \times [3] = [1] + [2]+[3] \,.
\end{equation}
The fusion characters read
\begin{equation}
  \begin{aligned}
    & \lambda_1=(1,2\cos\tfrac\pi7,1+2\cos\tfrac{2\pi}7) \,, \\
    & \lambda_2=(1,2\cos\tfrac{3\pi}7,1+2\cos\tfrac{6\pi}7) \,, \\
    & \lambda_3=(1,2\cos\tfrac{5\pi}7,1+2\cos\tfrac{10\pi}7) \,.
  \end{aligned}
\end{equation}
The only category automorphism is the identity. The category is braided.

\paragraph{Face model with $a=3$.}
The adjacency matrix $G=\Nh(3)$ corresponds to the graph
\begin{equation}
  \mathfig{\includegraphics{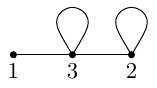}}
\end{equation}
The quantum dimension $d(3)=1+2\cos\tfrac{2\pi}7$ is greater than two.
Setting $v=1$ on every face, we obtain the Boltzmann weights
\begin{equation}
  \begin{array}{cccccccc}
    \sideset{_{\ds 1}^{\ds 3}}{_{\ds 3}^{\ds 1}}{\face{face.pdf}}
    & \sideset{_{\ds 2}^{\ds 3}}{_{\ds 3}^{\ds 2}}{\face{face.pdf}}
    & \sideset{_{\ds 2}^{\ds 2}}{_{\ds 2}^{\ds 2}}{\face{face.pdf}} 
    & \sideset{_{\ds 3}^{\ds 3}}{_{\ds 3}^{\ds 3}}{\face{face.pdf}}  \\[20pt]
    \frac{d(1)+d(3)}{\sqrt{d(1)d(3)}} & \frac{d(2)+d(3)}{\sqrt{d(2)d(3)}} & 2 & 2 \\
    \ \\
    \sideset{_{\ds 1}^{\ds 3}}{_{\ds 3}^{\ds 2}}{\face{face.pdf}} 
    & \sideset{_{\ds 1}^{\ds 3}}{_{\ds 3}^{\ds 3}}{\face{face.pdf}} 
    & \sideset{_{\ds 3}^{\ds 2}}{_{\ds 2}^{\ds 2}}{\face{face.pdf}}   
    & \sideset{_{\ds 2}^{\ds 3}}{_{\ds 3}^{\ds 3}}{\face{face.pdf}}   \\[20pt]
    \left[\frac{d(1)d(2)}{d(3)^2}\right]^{1/4} & \left[\frac{d(1)}{d(3)}\right]^{1/4}
    & \left[\frac{d(3)}{d(2)}\right]^{1/4} & \left[\frac{d(2)}{d(3)}\right]^{1/4}
  \end{array}
\end{equation}
and the interactions are invariant under a $\pi/2$ rotation, which yields the weights of the other face configurations.

\paragraph{Analysis of the lattice model.}
Applying the method presented in Sec.~\ref{sec:decomp} for fixed boundary conditions, we obtain the decompositions
\begin{equation}
  \begin{aligned}
    & \V_{11}^{\rm even} \equiv I_0 \oplus I_{2} \oplus 4I_3 \oplus 8 I_4 \oplus 24 I_5 \oplus \dots \\    
    & \V_{12}^{\rm even} \equiv I_1 \oplus 2 I_2\oplus 6I_3\oplus 16 I_4 \oplus 42 I_5 \oplus \dots \\
    & \V_{13}^{\rm even} \equiv I_1 \oplus I_{\frac32} \oplus 3I_2 \oplus 7I_3 \oplus 20 I_4 \oplus 53 I_5 \oplus \dots \\
    & \V_{22}^{\rm even} \equiv I_0 \oplus I_1 \oplus 4I_2 \oplus 11 I_3 \oplus 28 I_4 \oplus 77 I_5 \oplus \dots \\
    & \V_{23}^{\rm even} \equiv 2I_1  \oplus 5 I_2 \oplus 13I_3 \oplus 36I_4 \oplus 95 I_5 \oplus\dots \\
    & \V_{33}^{\rm even} \equiv I_0 \oplus 2I_1\oplus 6 I_2 \oplus 17I_3\oplus 44I_4 \oplus 119 I_5 \oplus\dots
  \end{aligned}
\end{equation}
and
\begin{equation}
  \begin{aligned}
    & \V_{11}^{\rm odd} \equiv I_{\frac32} \oplus 2I_{\frac52} \oplus 5I_{\frac72}\oplus 15 I_{9/2} \oplus \dots \\
    & \V_{12}^{\rm odd} \equiv 2 I_{\frac32} \oplus 3 I_{\frac52} \oplus 10 I_{\frac72} \oplus 26 I_{\frac92} \oplus \dots \\
    & \V_{13}^{\rm odd} \equiv I_{\frac12} \oplus 5I_{\frac52} \oplus 12 I_{\frac72} \oplus 32 I_{\frac92} \oplus \dots \\
    & \V_{22}^{\rm odd} \equiv I_{\frac12}\oplus 2I_{\frac32}\oplus 7I_{\frac52} \oplus 17 I_{\frac72} \oplus 47I_{\frac92} \oplus\dots \\
    & \V_{23}^{\rm odd} \equiv I_{\frac12} \oplus 3I_{\frac32}\oplus 8I_{\frac52}\oplus 22I_{\frac72}\oplus 58I_{\frac92}  \oplus \dots \\
    & \V_{33}^{\rm odd} \equiv I_{\frac12}\oplus 4I_{\frac32}\oplus  10 I_{\frac52}\oplus 27I_{\frac72}\oplus 73I_{\frac92}\oplus\dots
  \end{aligned}
\end{equation}
For periodic boundary conditions, we obtain
\begin{equation}
  \begin{aligned}
    \Vh^{\rm even} &\equiv
    I_{0,z_1} \oplus I_{0,z_2} \oplus I_{0,z_3} \oplus I_{1,1}
    \oplus 3I_{2,1} \oplus I_{2,-1} \oplus 2(I_{2,i}\oplus I_{2,-i})  \\
    &\qquad \oplus 5I_{3,1} \oplus 2I_{3,-1} \oplus 3(I_{3,e^{i\pi/3}}
    \oplus I_{3,e^{-i\pi/3}}) \oplus 4(I_{3,e^{2i\pi/3}}\oplus I_{3,e^{-2i\pi/3}}) \oplus \dots \\
    \Vh^{\rm odd} &\equiv 2I_{\frac12,1} \oplus 3I_{\frac32,1} \oplus I_{\frac32,e^{2i\pi/3}}\oplus I_{\frac32,e^{-2i\pi/3}}  \\
    &\qquad \oplus 4I_{\frac52,1} \oplus 2(I_{\frac52,e^{2i\pi/5}}
    \oplus I_{\frac52,e^{-2i\pi/5}}\oplus I_{\frac52,e^{4i\pi/5}}\oplus I_{\frac52,e^{-4i\pi/5}}) \oplus \dots 
  \end{aligned}
\end{equation}
where $z_i$ is a solution of $z_i+z_i^{-1}=\lambda_i(3)$. Note that $I_{0,z_1}=\I_{11}$ is the module containing the ground state.
\medskip

Since $d(3)>2$, the model is not critical, i.e. it has a finite correlation length in the scaling limit.

\section{Modular invariance}

\subsection{The $A_n$ model}
\label{sec:An}

\paragraph{Fusion category data.}
We consider the $A_n$ category, where $n$ is an integer such that $n\geq 3$. The simple objects are $\{1,2,\dots,n\}$, and the fusion rules read
\begin{equation} \label{eq:Nabc.An}
  N_{ab}^c = \begin{cases}
    1 & \text{if } \quad \begin{aligned}
      & a+b\geq c+1 \,, \ b+c\geq a+1 \,, \ c+a\geq b+1  \\
      & a+b+c\leq 2n+1 \,, \ a+b+c \in 2\Zbb+1 \,,
    \end{aligned} \\
    0 & \text{otherwise.}
  \end{cases}
\end{equation}
The object $1$ is the identity, i.e. $[1]\times [a]=[a]$, whereas the object $n$ acts as the $\Zbb_2$ permutation $[n]\times [a] = [n+1-a]$.
The only category automorphism is the identity.
\medskip

Using the relation \eqref{eq:Na.Nb}, one obtains
\begin{equation}
  \wh{N}(2) \, \wh{N}(b) =
  \begin{cases}
    \wh{N}(b-1) + \wh{N}(b+1) & \text{if } 2\leq b\leq n-1 \\
    \wh{N}(2) & \text{if } b=1 \\
    \wh{N}(n-1) & \text{if } b=n-1 \,.
  \end{cases}
\end{equation}
Hence, the fusion matrices for $b=1,\dots,n$ read $\wh{N}(b)=U_{b-1}[\wh{N}(2)/2]$, where $U_m$ is the $m$-th Chebyshev polynomial of the second kind.
The fusion characters are $\lambda_1,\dots,\lambda_n$, with
\begin{equation}
  \lambda_j(m)= \frac{\sin \frac{mj\pi}{h}}{\sin \frac{j\pi}{h}} =U_{m-1}\left[\frac{\lambda_j(2)}{2}\right]\,,
  \qquad h=n+1 \,.
\end{equation}
In particular, the quantum dimensions read
\begin{equation}
  d(m)= \lambda_1(m)=\frac{\sin \frac{m\pi}{h}}{\sin \frac{\pi}{h}} =U_{m-1}\left[\frac{d(2)}{2}\right]\,.
\end{equation}
The $F$-symbols are given in \cite{KL94} -- see also \cite{Guide}.
The category is braided, and one has $R_{22} = q^{1/2} \id_{22} + q^{-1/2} \psi_1^{22}\circ \psi_{22}^1$.
The $S$-matrix reads
\begin{equation} \label{eq:S.An}
  S_{ab} = S_{ba} = \frac{\lambda_a(b)}{|\lambda_a|}
  = \frac{\lambda_b(a)}{|\lambda_b|} = \sqrt{\frac 2h} \sin\frac{\pi ab}{h} \,,
  \qquad h=n+1 \,,
\end{equation}
and the twist operator is
\begin{equation} \label{eq:T.An}
  T=\mathrm{diag}(\theta_s)_{s=1,\dots,h-1} \,,
  \qquad \theta_s= e^{i\pi(1-s^2)/(2h)} \,.
\end{equation}
The operators $S$ and $T$ form a representation of the modular group. More specifically, we have $S^2=\id$ and $(ST)^3=\Theta\,\id$, where $\Theta$ is a complex number of modulus one.

\paragraph{Face model.} We set $a=2$.
The adjacency matrix for $\wh{N}(2)$ corresponds to the Dynkin diagram
\begin{equation} \label{eq:An}
  A_n = \quad \mathfig{\includegraphics{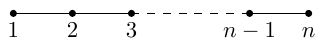}}
\end{equation}
The loop weight is then $\beta=d(2)=2\cos\tfrac{\pi}{n+1}$, corresponding to $h=n+1$ and $q=\exp(\tfrac{i\pi n}{n+1})$. Let us set $v=1$ on every face
The resulting face model is the Andrews-Baxter-Forrester model at the critical point, with Boltzmann weights
\begin{equation}
  \begin{array}{cc}
    \sideset{_{\ds \quad x}^{\ds x\!+\!1}}{_{\ds x\!+\!1}^{\ds x}}{\face{face.pdf}}
    & \sideset{_{\ds \quad x}^{\ds x\!+\!1}}{_{\ds x\!-\!1}^{\ds x}}{\face{face.pdf}} \\[20pt]
    \frac{d(x)+d(x+1)}{\sqrt{d(x)d(x+1)}}
    & \left[\frac{d(x+1)d(x-1)}{d(x)^2}\right]^{1/4}
  \end{array}
\end{equation}
and the other face weights are obtained by $\pi/2$ rotations.

\paragraph{Decomposition of the space of states.}
We consider the case of twisted periodic boundary conditions, and we use the short-hand notation $\Vh_m:=\Vh_{m,\id}$ for any $m=1,\dots,n$. Due to the bipartition of the adjacency graph \eqref{eq:An}, the space $\Vh_m(N)$ is non-zero only if $(m+N)$ is odd. We shall then argue that the following decomposition holds
\begin{equation} \label{eq:decomp.Vm.An}
  \Vh_m(N) \equiv
  \bigoplus_{k=-\frac{m-1}{2}}^{\frac{m-1}{2}} \bigoplus_{\ell=\frac{m+1}{2}}^{h-\frac{m+1}{2}}  \I_{\ell+k,\ell-k}(N) \,,
\end{equation}
where $\I_{\ell+k,\ell-k}:=I_{k,(-1)^{\ell+k}q^\ell}$ as in \eqref{eq:I}.
\medskip

Let us first show that the dimensions of the LHS and RHS of \eqref{eq:decomp.Vm.An} agree. For the LHS, we use the dimension formula \eqref{eq:dim.VmK}, which can be written
\begin{equation}
  \dim \wh{\cal V}_m(N)=
  \sum_{k=0,\frac 12,\dots,h-\frac{1}{2}}  \alpha_k(m)\, D_k(N) \,,
  \qquad \alpha_k(m) = \Tr \left[\Nh(m)T_{2k}\left(\frac{\Nh(2)}{2}\right) \right] \,,
\end{equation}
where $T_{2k}$ denotes the $2k$-th Chebyshev polynomial of the first kind.
Using simple trigonometric calculations, we obtain for any $k=0,\tfrac12,\dots,h-\tfrac12$
\begin{equation}
  \alpha_k(1) = h\delta_{k,0}-1 \,,
  \qquad \alpha_k(2) = h(\delta_{k,\frac12}+\delta_{k,h-\frac12})-2 \,,
\end{equation}
and
\begin{equation}
  \alpha_k(m+1)-\alpha_k(m-1)
  = h(\delta_{k,\frac m2}+\delta_{k,h-\frac m2})-2 \qquad \text{if } m\geq 2 \,.
\end{equation}
For $m=1$ and $m=2$, this yields
\begin{equation} \label{eq:dim.Vm.An.ini}
  \begin{aligned}
    & \dim \Vh_1(N)= (h-1)D_0(N) -\sum_{\ell=1}^{h-1} D_s(N)
    = \sum_{\ell=1}^{h-1} \dim \I_{\ell\ell}(N) \,, \\
    & \dim \Vh_2(N)= 2(h-2)D_{1/2}(N) - 2\sum_{\ell=3/2}^{h-3/2} D_s(N)
    = \sum_{k=\pm 1/2}\sum_{\ell=3/2}^{h-3/2} \dim \I_{\ell+k,\ell-k}(N) \,.
  \end{aligned}
\end{equation}
For $m\geq 2$ we get the recursion relation
\begin{equation} \label{eq:rec1}
  \dim \Vh_{m+1}(N)-\dim \Vh_{m-1}(N)
  = 2h \,D_{m/2}(N) -2\sum_{k=0,\tfrac 12,\dots,h-\tfrac12} D_k(N) \,.
\end{equation}
On the other hand, the dimension of the RHS of \eqref{eq:decomp.Vm.An} reads
\begin{alignat}{1}
  \rho_m(N):=\sum_{k=-\frac{m-1}{2}}^{\frac{m-1}{2}}
  \sum_{\ell=\frac{m+1}{2}}^{h-\frac{m+1}{2}} \dim \I_{\ell+k,\ell-k}(N) 
  &= \sum_{k=-\frac{m-1}{2}}^{\frac{m-1}{2}}
  \sum_{\ell=\frac{m+1}{2}}^{h-\frac{m+1}{2}} [D_k(N)-D_\ell(N)] \nn \\
  &= (h-m) \sum_{k=-\frac{m-1}{2}}^{\frac{m-1}{2}} D_k(N)
  -m \sum_{\ell=\frac{m+1}{2}}^{h-\frac{m+1}{2}} D_\ell(N)
\end{alignat}
After some simple algebra, using the symmetries $D_k=D_{-k}=D_{h+k}$, we find for any $m\geq 2$
\begin{equation}
  \rho_{m+1}(N) - \rho_{m-1}(N) = 2h \,D_{m/2}(N) -2\sum_{k=0,\tfrac 12,\dots,h-\tfrac12} D_k(N) \,.
\end{equation}
Comparing with \eqref{eq:rec1}, we see that $\dim \wh{\cal V}_m(N)$ and $\rho_m(N)$ obey the same recursion. Since their initial values \eqref{eq:dim.Vm.An.ini} at $m=1,2$ also coincide, this proves that
\begin{equation} \label{eq:dim.Vm.An}
  \dim \Vh_m(N) =
  \sum_{k=-\frac{m-1}{2}}^{\frac{m-1}{2}} \sum_{\ell=\frac{m+1}{2}}^{h-\frac{m+1}{2}}  \dim\I_{\ell+k,\ell-k}(N)
\end{equation}
for any $m=1,2,\dots,n$.
\medskip

As a next step to establish the decomposition \eqref{eq:decomp.Vm.An}, we need to compute the eigenvalues of $f$ (resp. $\Omega$) on the space $\Vh_m(0)$ [resp. the spaces $\Vh_m(1),\Vh_m(2),\dots,\Vh_m(m-1)$]. Note that these operators are given explicitly in \eqref{eq:Omega.x} in terms of the $F$-symbols.
\medskip

For $m=1$, recall that $f$ acts as $\wh{N}(2)$ on $\Vh(0)$. Its eigenvalues read
\begin{equation}
  \lambda_1(2), \lambda_2(2), \dots, \lambda_n(2)
  = 2\cos\left(\frac{\pi}{h}\right), 2\cos\left(\frac{2\pi}{h}\right), \dots , 2\cos\left(\frac{n\pi}{h}\right)  \,,
\end{equation}
and hence its eigenvectors are insertion states of type $(0,-q),(0,q^2),\dots,(0,(-q)^n)$, since $-q=e^{-i\pi/h}$. Hence, the argument of Sec.~\ref{sec:decomp} yields \eqref{eq:decomp.Vm.An} for $\Vh(N)$.
\medskip

For $m=3,5,7\dots$ as argued in Sec.~\ref{sec:decomp}, the eigenvalues of $f$ on $\Vh_m(0)$ are all of the form $\lambda_\ell(2)=\cos(\pi \ell/h)$ with $1\leq \ell\leq h-1$. Moreover, Eq.~\eqref{eq:dim.Vm.An} yields $\dim \Vh_m(0)=h-m$. It is a non-trivial task to determine the spectrum of $f$ on $\Vh_m(0)$ among the possible values of $\ell$, and we do this using computer algebra for several values of the parameters $(h,m)$, which amounts to diagonalising the $(h-m)\times(h-m)$ matrix representing $f$ \eqref{eq:Omega.x} on $\Vh_m(0)$. We find that the spectrum reads
\begin{equation}
  \left\{ 2\cos \frac{\pi \ell}{h} \,, \quad \ell=\frac{m+1}2,\dots,h-\frac{m+1}2 \right\} \,.
\end{equation}
Similarly, Eq.~\eqref{eq:dim.Vm.An} yields $\dim \Vh_m''(2k)=2(h-m)$ for $k=1,2,3\dots$ where $\Vh_m''$ is defined in \eqref{eq:VmK'}. Using computer algebra for several values of $(h,m,k)$, we obtain the spectrum of $\Omega$ \eqref{eq:Omega.x} on $\Vh_m''(2k)$:
\begin{equation}
  \left\{ (-1)^{\ell+k}q^\ell \,, \quad \ell= \pm\frac{m+1}2,\dots,\pm \left(h-\frac{m+1}2 \right) \right\} \,.
\end{equation}
These spectra of $f$ and $\Omega$ together with the dimension formula \eqref{eq:dim.Vm.An} yield \eqref{eq:decomp.Vm.An} for $\Vh_m(N)$.
\medskip

For $m=2,4,6\dots$ since $\Vh_m(N)=0$ if $N$ is even, we only need to consider the spectrum of $\Omega$ on $\Vh_m(1),\Vh_m''(3),\Vh_m''(5)\dots$ We perform a similar analysis using computer algebra for several values of $(h,m,k)$, and we find the same form for the spectrum of $\Omega$, which confirms \eqref{eq:decomp.Vm.An}.

\paragraph{Topological operators.}
Since $R_{22}$ and $\Rb_{22}$ are equal to the TL braid operators, the operators $Y_2$ and $\Yb_{\!\!2}$ on $\Vh_m(N)$ coincide respectively with the braid matrices $F$ and $\Fb$, and they have eigenvalues given by \eqref{eq:F.I} on the submodule $\I_{\ell+k,\ell-k}(N)$. Hence we have
\begin{equation}
  Y_2 = 2\cos\frac{\pi(\ell+k)}{h} \,\id \,,
  \qquad \Yb_{\!\!2} = 2(-1)^N\cos\frac{\pi(\ell-k)}{h} \,\id
  \qquad \text{on } \I_{\ell+k,\ell-k}(N) \,.
\end{equation}
From the fusion relations, one has
\begin{equation}
  Y_p = U_{p-1}(Y_2/2) \,, \qquad \Yb_{\!\!p} = U_{p-1}(\Yb_{\!\!2}/2) \,,
\end{equation}
where $U_{p-1}$ is the Chebyshev polynomial of the second kind.
This yields
\begin{equation} \label{eq:Y.I}
  Y_p = \lambda_{\ell+k}(p) \,\id \,,
  \qquad \Yb_{\!\!p}=(-1)^{(p+1)N}\lambda_{\ell-k}(p) \,\id
  \qquad \text{on } \I_{\ell+k,\ell-k}(N) \,.
\end{equation}
Note that $\lambda_{h-j}(p)=(-1)^{p+1}\lambda_j(p)$. 

\subsection{Temperley-Lieb modular characters}
\label{sec:chi}

Let us define the modular characters associated to a $\TL^a$ module $A$. Let $M,N$ be two non-negative integers, and $\mu$ an element of $\TL(M+N)$. We represent $\mu$ in a rectangle with $M$ nodes on the vertical sides, and $N$ nodes on the horizontal sides
\begin{center}
  \includegraphics{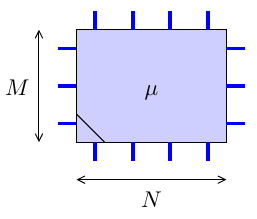}
\end{center}
The nodes on the left and bottom sides are interpreted as the incoming nodes, and those on the top and right sides are the outgoing nodes.
For instance, $(\id+v e_1)$ corresponds to a single face, $(\id+v e_2)(\id+v e_1)$ corresponds to a system of size $2\times 1$, etc.
For any $\TL^a$ module $A$ and any $\mu\in\TL(M+N)$ we define the characters
\begin{equation}
  \begin{aligned}
    & \chi_A(\mu,M,N):=
    \Tr_{A(N)}\left[ c_{N+1}c_{N+2}\dots c_{N+M} \cdot \mu \cdot (c_0^\dag)^M \right] \,, \\
    & \wt\chi_A(\mu,M,N) :=
    \Tr_{A(M)}\left[ c_1c_2\dots c_N \cdot \mu \cdot (c_0^\dag)^N \right] \,.
  \end{aligned}
\end{equation}
In the following, to lighten the notation, we shall write $\chi_A(\mu)$ and $\wt\chi_A(\mu)$ instead of $\chi_A(\mu,M,N)$ and $\wt\chi_A(\mu,M,N)$.
Note that $c_{N+1}c_{N+2}\dots c_{N+M} \mu (c_0^\dag)^M$ [resp. $c_1c_2\dots c_N \mu (c_0^\dag)^N$] is an element of $\TL^a(N)$ [resp. $\TL^a(M)$], obtained by connecting together the left and right (resp. top and bottom) sides of $\mu$. For any $\mu\in\TL(M+N)$, we define the diagram $\wt\mu \in \TL(M+N)$ as
\begin{equation}
  \wt\mu := c_{N+M+1} c_{N+M+2} \dots c_{N+2M} \cdot (\id_M \otimes \mu \otimes \id_M) \cdot c^\dag_M c^\dag_{M-1} \dots c^\dag_1 \,.
\end{equation}
With this definition, we obtain $\wt\chi_A(\mu,M,N)=\chi_A(\wt\mu,N,M)$.
Thus, graphically, for instance with $N=4$ and $M=3$, we have
\begin{equation}
  \chi_A(\mu)=
  \Tr_{A(N)}\left[ \mathfig{\includegraphics{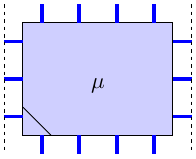}} \right] \,,
  \qquad
  \chi_A(\wt\mu)=
  \Tr_{A(M)}\left[ \mathfig{\includegraphics{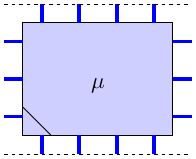}} \right] \,,
\end{equation}
where the dotted lines indicate periodic boundary conditions.

\medskip

In the following, we consider the algebra $\TL^a$ of loop weight $\beta=-q-q^{-1}$, with  $q=e^{i\pi(h-1)/h}$, where $h$ is a positive integer. The simple modules are of the form
\begin{equation}
  \I_{\ell+k,\ell-k} = I_{k,(-1)^{\ell+k}q^\ell} \,,
  \qquad |k|\leq \frac h2-1 \,, \quad 1\leq \ell\leq h-1\,, \quad \ell+k\in\Zbb \,,
\end{equation}
and we use the short-hand notation $\chi_{s,\sbar}(\mu):=\chi_{\I_{s,\sbar}}(\mu)$ for the TL modular characters.
\medskip

In the $A_n$ model, since the twisted boundary condition $(m,\id)$ and the topological operators $Y_m,\Yb_{\!\!m}$ consist in inserting the weights \eqref{eq:seam} respectively on a vertical and horizontal closed path, we obtain the relation
\begin{equation}
  \chi_{\Vh_m}(Y_p \cdot \mu) = \chi_{\Vh_p}(\Yb_{\!\!m}\cdot \wt\mu) \,,
\end{equation}
for any $\mu\in\TL(M+N)$ and any pair of integers $(m,p)$ with $1\leq m,p\leq n$.
We note that the decomposition \eqref{eq:decomp.Vm.An} can be written as
\begin{equation}
  \Vh_m(N) \equiv
  \bigoplus_{\smallarrayy{|k|\leq \frac h2-1}{1\leq \ell\leq h-1}{\ell+k\in\Zbb}} N_{\ell+k,\ell-k}^m \, \I_{\ell+k,\ell-k}(N) \,.
\end{equation}
Hence, using \eqref{eq:Y.I} we obtain
\begin{equation}
  \sum_{\ell,k} N_{\ell+k,\ell-k}^m \, \lambda_{\ell+k}(p) \, \chi_{\ell+k,\ell-k}(\mu)
  = (-1)^{MN} \sum_{i,j} N_{i+j,i-j}^p \, \lambda_{i-j}(m) \, \chi_{i+j,i-j}(\wt\mu) \,.
\end{equation}
We fix two integers $s,\tb$ such that $1\leq s,\tb \leq n$, we multiply the above relation by
the factor $$\frac{\lambda_s(p)\lambda_{\tb}(m)}{|\lambda_s|^2 |\lambda_{\tb}|^2}$$
and we sum over $m$ and $p$. This yields
\begin{equation}
  \sum_{\ell+k=s} \frac{\lambda_{\tb}(\ell-k)}{|\lambda_{\tb}|} \, \chi_{\ell+k,\ell-k}(\mu)
  = (-1)^{MN} \sum_{i-j=\tb} \frac{\lambda_s(i+j)}{|\lambda_s|} \, \chi_{i+j,i-j}(\wt\mu) \,.
\end{equation}
where we have used \eqref{eq:S.An}, \eqref{eq:fusion.lambda} and \eqref{eq:lambdai.lambdaj}.
We now fix an integer $\sbar$ such that $1\leq \sbar \leq n$, we multiply by $\lambda_{\sbar}(\tb)/|\lambda_{\sbar}|$, and we sum over $\tb$, which gives
\begin{equation}
  \chi_{s,\sbar}(\mu)
  = (-1)^{MN} \sum_{1\leq t,\tb\leq n} \frac{\lambda_s(t)\lambda_{\sbar}(\tb)}{|\lambda_s||\lambda_{\sbar}|} \, \chi_{t,\tb}(\wt\mu) \,.
\end{equation}
Recall that $\I_{s,\sbar}(N)$ is nonzero only if $(N+s+\sbar)$ is even. Hence the above relation can be written as
\begin{equation} \label{eq:modular}
  \chi_{s,\sbar}(\mu)
  =  \sum_{t,\tb=1}^{h-1}
  \mathcal{S}_{(s,\sbar),(t,\tb)} \, \chi_{t,\tb}(\wt\mu) \,,
  \qquad \mathcal{S}_{(s,\sbar),(t,\tb)}=(-1)^{(s+\sbar)(t+\tb)}S_{st}\, S_{\sbar\tb} \,,
\end{equation}
where $(S_{st})$ is the $S$-matrix of the $A_n$ category, given in \eqref{eq:S.An}. Note that a result similar to \eqref{eq:modular} was already derived in \cite{PasquierSaleur90}.
\medskip

The twist operator $\mathcal{T}=\Omega^N$ has eigenvalue $[(-1)^s q^{(s+\sbar)/2}]^{s-\sbar}$ on $\mathcal{I}_{s,\sbar}(N)$. Comparing with \eqref{eq:T.An}, we obtain
\begin{equation}
  \chi_{s,\sbar} \left(\Omega^N\mu \right)
  =  \theta_{s,\sbar} \, \chi_{s,\sbar}(\mu) \,,
  \qquad \theta_{s,\sbar}=e^{-i\pi(s+\sbar)^2/2} \, \theta_s \, \theta_{\sbar}^* \,.
\end{equation}
After some simple algebra, using the properties of $S$ and $T$, we obtain
\begin{equation}
  \mathcal{S}^2 = \id \,, \qquad (\mathcal{ST})^3 = \id \,,
\end{equation}
where $\mathcal{T}=\mathrm{diag}(\theta_{s,\sbar})$. Hence, the modular characters $\chi_{s,\sbar}$ form a representation of the modular group.
\medskip

Applying the Verlinde formula to the $\mathcal{S}$-matrix, we obtain the fusion numbers
\begin{equation} \label{eq:N.TL}
  \mathcal{N}_{(s,\sbar),(t,\tb)}^{(u,\ub)}
  := \sum_{1\leq v,\vb \leq h-1} \frac{\mathcal{S}_{(s,\sbar),(v,\vb)}\mathcal{S}_{(t,\tb),(v,\vb)}\mathcal{S}_{(u,\ub),(v,\vb)}}{\mathcal{S}_{(1,1),(v,\vb)}} = N_{st}^u \, N_{\sbar\tb}^{\ub} \,,
\end{equation}
where $N_{ab}^c$ is given in \eqref{eq:Nabc.An}.
This suggests that the fusion rules for the simple modules under the $\TL^a$ fusion product \cite{PTLFunctor} are given by
\begin{equation}
  \I_{s,\sbar} \otimes \I_{t,\tb} \equiv
  \sum_{1\leq u,\ub \leq h-1} N_{st}^u \, N_{\sbar\tb}^{\ub} \, \I_{u,\ub} \,.
\end{equation}

\paragraph{Scaling limit.}
The monodromy matrix $L_N \in \TL(N+1)$ and the row-to-row transfer matrix $t_N\in \TL^a(N)$ are defined as
\begin{equation}
  L_N := (\id+v e_N)\dots(\id+v e_2)(\id+v e_1) \,,
  \qquad t_N := c_{N+1} \cdot L_N \cdot c^\dag_0 \,,
\end{equation}
where $v$ is the local coupling constant as in \eqref{eq:face}.
For instance, in any face model defined as in Sec.~\ref{sec:def}, the partition function on the $N\times M$ lattice with periodic boundary conditions is given by
\begin{equation}
  Z_{M,N} = \Tr_{\Vh(N)}\left[(t_N)^M \right]
  = \chi_{\Vh}\left[
    (L_N\otimes \id_{M-1})(\id_1\otimes L_N\otimes \id_{M-2})\dots(\id_{M-1}\otimes L_N)
    \right] \,.
\end{equation}
The dominant eigenvalue of $t_N$ on $\Vh(N)$ typically scales as $\exp(-Nf_\infty)$, where $f_\infty$ is the free energy density per face, and the scaling corrections to this behaviour are incoded in the CFT.

To make contact with Virasoro characters, one considers the following collection of TL characters. For any $\tau\in \Cbb$ such that $\Im\,\tau>0$ and any non-negative integer $N$, we write
\begin{equation}
  M_1(N,\tau):= \lfloor \Re(N\tau)/2\rfloor \,,
  \qquad M_2(N,\tau):= \lfloor \Im(N\tau)/2\rfloor \,,
\end{equation}
and we define, for any $\TL^a$ module $A$, and any $m_1,m_2\in\{0,1\}$
\begin{equation}
  \chi_A(\tau,m_1,m_2):= \lim_{N\to \infty}
  \Tr_{A(N)}\left[ \Omega^{2M_1(N,\tau)+m_1}\cdot \left(e^{Nf_\infty}t_N \right)^{2M_2(N,\tau)+m_2} \right] \,,
\end{equation}
where $\Omega$ is the shift operator, and the parameter $v$ is set to one.
From the scaling \eqref{eq:scaling.T.Om}, one has for $A=\I_{s,\sbar}$
\begin{equation}
  \chi_{s,\sbar}(\tau,m_1,m_2)= \sum_{r=1}^{h'-1} (-1)^{(r+s)(m_1+m_2)}
  \chi^{\rm Vir}_{r,s}(\tau)\, \chi^{\rm Vir}_{r,\sbar}(\tau)^*
\end{equation}
where $\chi^{\rm Vir}_{r,s}$ is the Virasoro character of the irreducible module with heighest weight $\Delta_{rs}$.
\medskip

For simplicity, we restrict to the case when $\tau$ is pure imaginary.
As $N\to\infty$, \eqref{eq:modular} yields
\begin{equation}
  \begin{aligned}
    \chi_{s,\sbar}(\tau,0,m)
    &= \sum_{r=1}^{h'-1} (-1)^{(r+s)m}
    \chi^{\rm Vir}_{r,s}(\tau)\, \chi^{\rm Vir}_{r,\sbar}(\tau)^* \\
    &= \sum_{\rho=1}^{h'-1} \sum_{\smallarray{t,\tb=1}{m+t+\tb \text{ even}}}^{h-1}
    (-1)^{(s+\sbar)(\rho+t+m)} S_{st}\, S_{\sbar\tb} \,\chi^{\rm Vir}_{\rho,t}(-1/\tau) \chi^{\rm Vir}_{\rho,\tb}(-1/\tau)^* \,,
  \end{aligned}
\end{equation}
which is consistent with the well-known expression of the $S$-matrix for Virasoro characters in the $\mathcal{M}(h,h')$ minimal CFT
\begin{equation}
  S^{\rm Vir}_{(r,s),(\rho,\sigma)}
  = \sqrt{\frac{8}{hh'}} (-1)^{(r+s)(\rho+\sigma)} \sin\frac{\pi r\rho}{h'}
  \sin\frac{\pi s\sigma}{h} \,.
\end{equation}

\section{Conclusion}
The decomposition of the space of states under the spectrum-generating algebra is a cornerstone for the symmetry analysis of any discrete or continuous model. In rational CFTs, the Hilbert space is usually determined, following \cite{CIZ87}or \cite{CardyBCFT89}, by consistency conditions on the modular partition functions. In the present work, for any face model with TL interactions associated to a fusion category, we have presented a systematic way of computing this decomposition from first principles. Moreover, applying this approach to the unitary series of RSOS models of \cite{ABF84}, we have obtained the modular transformation of TL characters, which is a fundamental result for the representation theory of the affine TL algebra.
\medskip

On the three specific examples of critical models we have considered, the topological operators of the form $Y_p$ and $Q_L$ generate the algebra of $\TL$ or $\TL^a$ endomorphisms, say $\mathrm{End}(\Vh)$ for periodic boundary conditions. In contrast, for face models with a loop weight $\beta \geq 2$ such as the psu(2)$_5$ face model, the dimension of $\mathrm{End}(\Vh)$ is infinite, whereas the operators $Y_p$ and $Q_L$ can only generate a finite-dimensional algebra. It would be interesting to find the concrete form of the other endomorphisms in such cases.
\medskip

Another perspective for future work is concerned with the formula \eqref{eq:N.TL} for the fusion numbers among simple $\TL^a$ modules $\I_{s\sbar}$. Indeed, the validity of the Verlinde relies on a set of conditions on the category of $\TL^a$ modules generated by the $\I_{s\sbar}$, which one would need to verify. Alternatively, one could think of calculating from first principles the decomposition of the fusion product $\I_{s\sbar} \otimes \I_{t,\tb}$, and check if it agrees with the fusion numbers \eqref{eq:N.TL}.

\section*{Acknowledgements}
The author wishes to thank Lea Bottini, Cl\'ement Delcamp, Beno\^{\i}t Dou\c{c}ot, Beno\^{\i}t Estienne, Eric Vernier and Jean-Bernard Zuber for insightful discussions, and Alexi Morin-Duchesne for collaborations on related work.

\appendix
\section*{Appendix}

\section{Fusion categories}
\label{app:fusion.cat}

\paragraph{Fusion category.} A fusion category $\C$ consists essentially in the following data \cite{Kitaev06}:
\begin{enumerate}
\item a finite set of simple objects $\{1,2,\dots,n\}$, equipped with some associative fusion rules
  \begin{equation} \label{eq:fusion.rules}
    a \otimes b = \bigoplus_{c=1}^n N_{ab}^c \, c \,,
  \end{equation}
  where $N_{ab}^c$ is a non-negative integer, and such that the object $1$ is the neutral element, i.e. $N_{1a}^b=N_{a1}^b=\delta_{ab}$,
\item for any triplet $a,b,c$ of simple objects, two vector spaces $V_{ab}^c$ and $V_c^{ab}$, encoding the morphisms $(a\otimes b) \to c$ and  $c\to (a\otimes b)$ respectively, with $\dim V_{ab}^c=\dim V_c^{ab}=N_{ab}^c$, and an antilinear involution $\dag$ between  $V_{ab}^c$ and $V_c^{ab}$,
\item an involution on simple objects $a \leftrightarrow \ab$, such that $N_{a\ab}^1=N_{\ab a}^1=1$ ($\ab$ is called the dual of $a$), and for each pair $(a,\ab)$, a pair of morphisms $\xi_a \in V_1^{a\ab}$ and $\eta_a \in V_1^{\ab a}$ such that
  $(\id_a \otimes \eta_a^\dag) \circ (\xi_a \otimes \id_a)=(\id_a \otimes \xi_a^\dag) \circ (\eta_a \otimes \id_a)= \id_a$ and $\xi_a^\dag \circ \xi_a=\eta_a^\dag \circ \eta_a := d(a)$ -- the number $d(a)$ is called the quantum dimension of $a$,
\item a collection of $F$-symbols $F_u^{abc}$ encoding the associativity isomorphism $\oplus_x (V_x^{ab}\otimes \id_c) \circ V_u^{xc} \to \oplus_y (\id_a \otimes V_y^{bc})\circ V_u^{ay}$,
  and obeying the pentagon equation.
\end{enumerate}
For simplicity, in the present work we make the following assumptions:
\begin{enumerate}
\item the fusion numbers are symmetric, i.e. $N_{ab}^c=N_{ba}^c$ for any $a,b,c$,
\item the fusion rules are multiplicity-free, i.e. $N_{ab}^c \in \{0,1\}$ for any $a,b,c$,
\item the category $\C$ is unitary, i.e. $d(a)>0$ for any $a$.
\end{enumerate}
Then every morphism space $V_{ab}^c$ (resp. $V_c^{ab}$) is spanned by a single state $\psi_{ab}^c$ [resp. $\psi_c^{ab}=(\psi_{ab}^c)^\dag$], and the states can be normalised so that\begin{equation}
\psi_{ab}^c\circ \psi_c^{ab} = \sqrt{\frac{d(a)d(b)}{d(c)}}\,\id_c \,,
\qquad \id_{ab} = \sum_{c=1}^n \sqrt{\frac{d(c)}{d(a)d(b)}} \, \psi_c^{ab} \circ \psi_{ab}^c \,.
\end{equation}
The quantum dimensions obey the relation
\begin{equation} \label{eq:da.db}
  d(a)d(b) = \sum_c N_{ab}^c \, d(c) \,.
\end{equation}
The associativity relations (also called $F$-moves) read
\begin{equation} \label{eq:F-move}
  (\psi_x^{ab}\otimes \id_c) \circ \psi_u^{xc}
  = \sum_{y=1}^n \left(F_u^{abc}\right)_{yx} \, (\id_a \otimes \psi_y^{bc}) \circ \psi_u^{ay} \,.
\end{equation}
Furthermore, to simplify the calculations, we assume that
\begin{equation}
  \eta_a =\xi_{\ab} \,,
  \qquad (\eta_b^\dag\otimes \id_c)\circ (\id_{\bb}\otimes \psi_a^{bc})=\psi_{\bb a}^c \,,
  \qquad (\id_b\otimes \xi_c) \circ (\psi_a^{bc} \otimes \id_{\cb}) = \psi_{a\cb}^b \,,
\end{equation}
i.e. we suppose that the Schur-Frobenius indicators and the index raising and lowering factors (denoted $\kappa_a,A_a^{bc},B_a^{bc}$ in \cite{Kitaev06}) are all equal to one -- these factors can be easily restored to recover the general case.
\medskip

We use the following conventions for graphical calculus:
\begin{equation}
  \psi_{ab}^c = \mathfig{\includegraphics{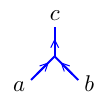}} 
  \qquad \psi_c^{ab} = \mathfig{\includegraphics{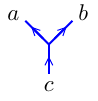}} 
  \qquad \xi_a=\psi_1^{a\ab} = \mathfig{\includegraphics{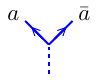}} 
  \qquad \xi_a^\dag=\psi_{a\ab}^1 = \mathfig{\includegraphics{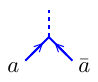}} 
\end{equation}

\paragraph{Fusion algebra.} For each simple object $a$, one defines the $n \times n$ adjacency matrix $[\Nh(a)]_{bc}:=N_{ab}^c$. The relation \eqref{eq:da.db} means that  $(d(1),\dots,d(n))$ is an eigenvector of $\Nh(a)$, with eigenvalue $d(a)$. By the Perron-Frobenius theorem, the quantum dimension $d(a)$ is thus equal to the dominant eigenvalue (i.e. the eigenvalue with maximal modulus) of $\Nh(a)$.
\medskip

Using the associativity of the fusion rules \eqref{eq:fusion.rules}, one shows that
\begin{equation} \label{eq:Na.Nb}
  \Nh(a) \cdot \Nh(b) = \sum_c N_{ab}^c \, \Nh(c) \,.
\end{equation}
In particular, the matrices $\Nh(a)$ commute with one another, and since $\Nh(\ab)=\Nh(a)^\dag$, they are normal matrices. Thus, they can be diagonalised simultaneously in an orthogonal basis $v_1,\dots,v_n$. Let $\lambda_j(a)$ be the eigenvalue of $\Nh(a)$ associated to $v_j$. The relation \eqref{eq:Na.Nb} yields
\begin{equation} \label{eq:fusion.lambda}
  \lambda_j(a) \, \lambda_j(b) = \sum_c N_{ab}^c \, \lambda_j(c) \,,
\end{equation}
and thus $(\lambda_j(1),\dots,\lambda_j(n))$ is an eigenvector of $\Nh(a)$, with eigenvalue $\lambda_j(a)$. Thus, we can choose the basis of common eigenvectors of the matrices $\Nh(a)$ as
\begin{equation}
  \lambda_j=(\lambda_j(1),\dots,\lambda_j(n)) \,, \qquad j=1,\dots,n \,.
\end{equation}
These vectors are called the fusion characters of the algebra \eqref{eq:fusion.rules}. They are orthogonal to one another:
\begin{equation} \label{eq:lambdai.lambdaj}
  \sum_{a=1}^n \lambda_i(a)^*\lambda_j(a) = \delta_{ij} \, |\lambda_j|^2 \,,
  \qquad
  |\lambda_j| := \sqrt{\sum_{a=1}^n |\lambda_j(a)|^2} \,.
\end{equation}
By convention, we set $\lambda_1=(d(1),\dots,d(n))$.
\medskip

An automorphism of $\C$ is a permutation $a\mapsto K(a)$ of the simple objects, such that
\begin{equation} \label{eq:automorphism}
  N_{K(a)K(b)}^{K(c)} = N_{ab}^c \qquad \text{and} \qquad
  \qquad \left(F_{K(u)}^{K(a)K(b)K(c)} \right)_{K(y)K(x)} = \left(F_u^{abc}\right)_{yx} \,.
\end{equation}
Let us introduce the matrix $\Kh$, with coefficients $\Kh_{ab}:=\delta_{a,K(b)}$. We then have $\Nh[K(a)]=\Kh \Nh(a) \Kh^{-1}$, and hence $d[K(a)]=d(a)$.

\paragraph{Braiding.} 
If the category $\C$ is braided, for each pair of simple objects $(a,b)$, there exists a pair of braid operators $R_{ab},\Rb_{ab}$ acting on morphisms as
\begin{equation}
  R_{ab} \circ \psi_c^{ab} = R_c^{ab} \, \psi_c^{ba}
  = \mathfig{\includegraphics{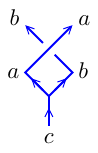}} 
 \qquad \Rb_{ab} \circ \psi_c^{ab} = \Rb_c^{ab} \, \psi_c^{ba}
  = \mathfig{\includegraphics{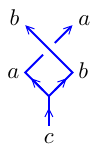}} 
\end{equation}
with $R_c^{ab},\Rb_c^{ab}\in\Cbb$, and obeying the hexagon equations.

\section{Temperley--Lieb representation theory}
\label{app:TL}

\subsection{Background material}

\paragraph{Temperley-Lieb diagrams and algebras.}

Consider the Temperley-Lieb algebra $\TL(N)$ with loop weight $\beta$, generated by
$e_1,\dots,e_{N-1}$, subject to \eqref{eq:TL.relations.ej}:
\begin{equation} 
  e_j^2 = \beta\, e_j \,,
  \qquad e_j e_{j\pm 1} e_j = e_j \,,
  \qquad e_j e_k=e_k e_j \quad \text{if } |j-k|>1 \,.
\end{equation}
The operator $e_j$ is depicted as the diagram
\begin{equation}
  e_j = \mathfig{\includegraphics{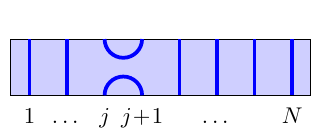}} \, \,.
\end{equation}
Following \cite{GL98}, we consider a generalisation of $\TL(N)$: we define $\TL(N,N')$ as the space of non-intersecting diagrams in a rectangle, connecting
$N$ nodes on the top boundary and $N'$ nodes on the bottom boundary.
Moreover, we define $\dag$ as the antilinear operator which maps any diagram $\lambda \in \TL(N,N')$ to its image $\lambda ^\dag\in \TL(N',N)$ under the reflection around the horizontal axis.
The diagram space $\TL(N-2,N)$ [resp. $\TL(N,N-2)$] is spanned by the elementary diagrams
$c_1,\dots,c_{N-1}$ (resp. $c^\dag_1,\dots,c^\dag_{N-1}$) given by
\begin{equation}
  c_j := \mathfig{\includegraphics{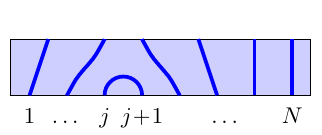}} \,,
  \qquad c^\dag_j := \mathfig{\includegraphics{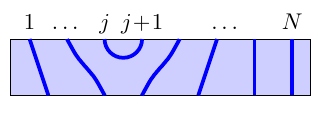}} \,.
\end{equation}
Here, by convention, the operators act from bottom to top.
The defining algebraic relations for these diagrams are
\begin{equation} \label{eq:TL.relations.cj}
  \begin{cases}
    c_j^{\phantom\dag}\, c_{j+1}^\dag = c_{j+1}^{\phantom\dag}\, c_j^\dag = \id  \\
    c_j^{\phantom\dag}\, c_j^\dag = \beta \, \id 
  \end{cases}
  \qquad \begin{cases}
    c_j c_k = c_k c_{j+2} \\
    c^\dag_k c^\dag_j = c^\dag_{j+2} c^\dag_k \\
    c^\dag_j c_k^{\phantom\dag} = c_k^{\phantom\dag} c^\dag_{j+2}
  \end{cases} \quad \text{if } k \leq j \,,
\end{equation}
and the generators of $\TL(N)$ are obtained as $e_j=c_j^\dag c_j^{\phantom\dag}$.
\medskip

Similarly, for periodic boundary conditions, one defines $\TL^a(N,N')$ as the space of non-intersecting diagrams in an annulus, connecting $N$ nodes on the top boundary and $N'$ nodes on the bottom boundary.
The elementary diagrams $c_0,\dots,c_{N-1}$ $\in \TL^a(N-2,N)$ and $c^\dag_0,\dots,c^\dag_{N-1}$ $\in \TL^a(N,N-2)$ obey the relations
\begin{equation} 
  \begin{aligned}
    &\begin{cases}
      c_j^{\phantom\dag}\, c_{j+1}^\dag = c_{j+1}^{\phantom\dag} \, c_j^\dag= \id  & \text{if } j \geq 1 \\ 
      c_j^{\phantom\dag}\, c_j^\dag = \beta \, \id  & \text{for any } j 
    \end{cases}
    && \begin{cases}
      c_j c_k = c_k c_{j+2} \\
      c^\dag_k c^\dag_j = c^\dag_{j+2} c^\dag_k \\
      c^\dag_j c_k^{\phantom\dag} = c_k^{\phantom\dag} c^\dag_{j+2}
    \end{cases} \quad \text{if } 1\leq k \leq j \\
    & \begin{cases}
      c_0c_j =c_{j-1}c_0   \\
      c_0^{\phantom\dag}c^\dag_j =c^\dag_{j-1}c_0^{\phantom\dag} \\
      c_j^{\phantom\dag}c_0^\dag = c_0^\dag c_{j-1}^{\phantom\dag} \\
      c^\dag_jc_0^\dag = c_0^\dag c^\dag_{j-1}
      \end{cases} \quad \text{if } 2 \leq j\leq N-2
    &\quad& \begin{cases}
      c_0c_{N-1}^\dag = c_1 c_0^\dag \\
      c_{N-1}^{\phantom\dag}c_0^\dag = c_0^{\phantom\dag}c_1^\dag \\
      c_0^{\phantom\dag}c_1^\dag \cdot c_1^{\phantom\dag} c_0^\dag = c_1^{\phantom\dag} c_0^\dag \cdot c_0^{\phantom\dag}c_1^\dag = \id 
    \end{cases}
  \end{aligned}
\end{equation}
where by convention, the rightmost $c_i$ (resp. $c_i^\dag$) in these expressions is in $\TL^a(N-2,N)$ [resp. $\TL^a(N,N-2)$].
One recovers the affine Temperley-Lieb algebra $\TL^a(N)$, by forming the generators
\begin{equation}
 \begin{aligned}
   & e_j=c^\dag_jc^{\phantom\dag}_j \,, \quad j=0,\dots,N-1 && \text{if } N \geq 2 \,, \\
   & \Omega=c_1c_0^\dag \,, \quad \Omega^\dag=c_0c_1^\dag \,, && \text{if } N \geq 1 \,, \\
   & f = c_1c_0^\dag = c_0c_1^\dag && \text{if } N=0 \,.
  \end{aligned}
\end{equation}
Here, the shift operators $\Omega,\Omega^\dag$ correspond to the diagrams
\begin{equation}
  \Omega = \mathfig{\includegraphics{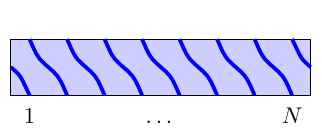}} \, \,,
  \qquad
  \Omega^\dag = \mathfig{\includegraphics{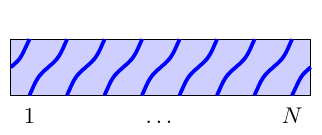}} \, \,,
\end{equation}
whereas $f$ inserts a non-trivial loop in the annulus with zero nodes on both boundaries.

\paragraph{Modules over $\TL$ and $\TL^a$.}
Following \cite{GL98}, one defines a $\TL$-module
as a collection
$$A=(A(0),A(1),A(2),\dots)$$
of modules over $\TL(0),\TL(1),\TL(2),\dots$ with an action $\TL(N,N'):A(N')\to A(N)$ obeying the relations \eqref{eq:TL.relations.cj}. Since $\TL(N,N')=0$ for any $N,N'$ with different parities, any $\TL$-module decomposes as
\begin{equation} \label{eq:M=Mev+Modd}
  A=A^{\rm even} \oplus A^{\rm odd}
  = (A(0),0,A(2),0,A(4),\dots) \oplus (0,A(1),0,A(3),0,A(5),\dots) \,.
\end{equation}
For any pair of modules $(A,B)$, a morphism $\phi:A \to B$ is a collection $(\phi_0,\phi_1,\phi_2,\dots)$ where $\phi_N$ is a linear map from $A(N)$ to $B(N)$, and
\begin{equation}
  \phi_N(\lambda \cdot u) = \lambda \cdot \phi_{N'}(u) \,,
\end{equation}
for any $\lambda \in \TL(N,N')$ and $u \in A(N')$.
The same definitions apply to $\TL^a$-modules.

\paragraph{Standard modules.}
The standard modules (also called Weyl modules) \cite{GL98} are an important class of $\TL$ or $\TL^a$ modules, which play a role analogous to heighest-weight modules of Lie algebras. Here, we review some basic facts on these modules -- see \cite{GL98} or \cite{PinetStAubin} for more details.
\medskip

We denote by $W_k$ the standard modules over $\TL$, with $2k\in \Zbb_{\geq 0}$. The basis states of $W_k(N)$ are connectivity states on $N$ nodes sitting on the boundary of a half-strip, with $2k$ legs and $(N/2-k)$ arches, and the $\TL$ diagrams act on them by connecting the loop segments (if $N-2k$ is odd or $N<2k$, then $W_k(N)=0$).
The dimension of $W_k(N)$ reads
\begin{equation}
  \dim W_k(N) = d_k(N)-d_{k+1}(N) \,,
\end{equation}
where for any $k \in \Zbb/2$
\begin{equation} \label{eq:def.dk}
  d_k(N) := \begin{cases}
    \left(\begin{array}{c} N \\ \frac N2-k \end{array}\right)
    & \text{if } |k|\leq \frac{N}{2} \text{ and } N+2k \text{ is even,} \\
    0 & \text{otherwise.}
  \end{cases}
\end{equation}
At $N=2k$, the module $W_k(2k)$ is one-dimensional, with a single state made of $2k$ legs, and denoted $u_k$. For any system size $N$, the action of $\TL(N,2k)$ on $u_k$ generates $W_k(N)$:
\begin{equation}
  W_k(N) = \TL(N,2k) \cdot u_k \,.
\end{equation}
The Gram form on $W_k$, denoted $\aaver{\,,\,}$, is defined as the unique bilinear form
$W_k(N) \times W_k(N) \to \Cbb$ which is invariant, i.e.
$\aaver{u,\lambda\,v} = \aaver{\lambda^t\,u, v}$
for any $u\in W_k(N), \ v \in W_k(N')$ and $\lambda \in \TL(N,N')$, and such that $\aaver{u_k,u_k}=1$.
Here the transposition is defined as the linear operator which maps any diagram $\lambda \in \TL(N,N')$ to its image $\lambda^t\in \TL(N',N)$ under the reflection around the horizontal axis (note that this is similar to $\lambda\mapsto \lambda^\dag$, except that the latter is antilinear).
\medskip

We denote by $W_{k,z}$ the standard modules over $\TL^a$, where $2k$ is the number of legs, and $z \in \Cbb^\times$ is the twist parameter. The space of states is similar to that of $W_k$, except that the arches connect nodes living on the boundary of a half-cylinder. The $\TL^a$ diagrams act by connecting the loop segments, and produce twist factors $z^{\pm 1}$ (resp. $z+z^{-1}$) as a leg crosses the boundary conditions (resp. as a non-contractible loop is formed).
In particular, the operator $\Omega^N$ acts as $z^{2k}\id$ in $W_{k,x}$.
The dimension of $W_{k,z}(N)$ reads
\begin{equation} \label{eq:dim.Wkz}
  \dim W_{k,z}(N)=d_k(N) \,,
\end{equation}
where $d_k(N)$ is given in \eqref{eq:def.dk}.
Like for $W_k$, the module $W_{k,z}(2k)$ is one-dimensional, and we denote by $u_k$ its unique basis state. We then have
\begin{equation}
  W_{k,z}(N)=\TL^a(N,2k)\cdot u_k \,.
\end{equation}
Similarly to $W_k$, one defines the Gram form on $W_{k,z}$ as the unique invariant bilinear form
$W_{k,1/z}(N) \times W_{k,z}(N) \to \Cbb$.

\paragraph{Simple modules.}
A simple module is a non-zero module which does not admit any non-trivial submodule.
Recall that a maximal submodule $A'$ of $A$ is a submodule such that $A/A'$ is simple, and the radical of $A$, denoted $\mathrm{rad}(A)$, is defined as the intersection of all its maximal submodules. A central result of \cite{GL98} is the following property on standard modules: the radical $\mathrm{rad}(W_k)$ [resp. $\mathrm{rad}(W_{k,z})$] is equal to the kernel of the Gram form in $W_k$ (resp. $W_{k,z}$), i.e.
\begin{equation}
  \begin{aligned}
    & \mathrm{rad}[W_k(N)] = \big\{ v \in W_k(N) \ | \ \aaver{u,v}=0 \text{ for any } u \in W_k(N) \big\} \,, \\
    & \mathrm{rad}[W_{k,z}(N)] = \big\{ v \in W_{k,z}(N) \ | \ \aaver{u,v}=0 \text{ for any } u \in W_{k,1/z}(N) \big\} \,.
  \end{aligned}
\end{equation}
Using this result, \cite{GL98} showed that the simple modules over $\TL$ and $\TL^a$ are all the modules of the form, respectively
\begin{equation}
  I_k:=W_k \big/ \mathrm{rad}(W_k) \qquad \text{with } 2k \in \Zbb
\end{equation}
and
\begin{equation}
  I_{k,z}:=W_{k,z} \big/ \mathrm{rad}(W_{k,z}) \qquad \text{with } 2k \in \Zbb \text{ and } z \in \Cbb^\times \,.
\end{equation}

To describe these modules more completely, it is convenient to parameterise the loop weight as
\begin{equation}
  \beta = -q-q^{-1} \,, \qquad q \in \Cbb^\times \,.
\end{equation}
We then distinguish two cases.
\begin{enumerate}
\item If $q$ is a root of unity, then we write $q=e^{i\pi h'/h}$ where $h,h'$ are two integers such that $0< h'<h$ and $\mathrm{gcd}(h,h')=1$. In this case, we shall restrict our attention to particular subsets of $\TL$ and $\TL^a$ simple modules. For $k=0,\tfrac 12,\dots,\tfrac h2-1$, one has $W_k \equiv [I_k \to I_{h-k-1}]$ \cite{RSA14}, and thus the dimensions of the corresponding simple modules read
  \begin{equation} \label{eq:dim.Ik}
    \dim I_k(N) = D_k(N) - D_{k+1}(N) \,,
  \end{equation}
  where
  \begin{equation} \label{eq:def.Dk}
    D_k(N):=\sum_{j\in \Zbb} d_{k+jh}(N) \,.
  \end{equation}
  For $k,m \in \Zbb/2$ and $\eps \in \{-1,+1\}$ with $0\leq k <m <h-k$ and $m-k\in \Zbb$, the module $W_{k,\eps q^m}$ has the double-ladder structure (see \cite{PinetStAubin})
  \begin{equation}
    W_{k,\eps q^m} \equiv \left[
      \mathfig{\includegraphics{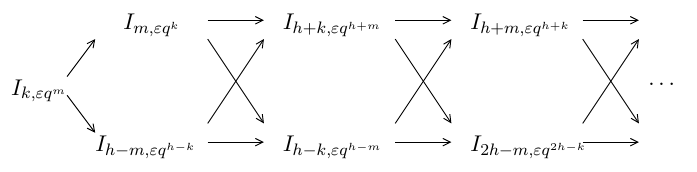}}
      \right]
  \end{equation}
  and similarly for $W_{k,\eps q^{-m}}$, up to the change $q\to q^{-1}$.
  The dimensions of the corresponding simple modules read
  \begin{equation}
    \dim I_{k,\eps q^{\pm m}}(N) = D_k(N) - D_m(N) \,.
  \end{equation}
  For the particular choice of sign $\eps=(-1)^{m+k}$, we shall denote these simple  modules as
  \begin{equation} \label{eq:I}
    I_{k,(-1)^{m+k} q^m} := \I_{m+k,m-k} \,.
  \end{equation}
  
\item If $q$ is not a root of unity then, for any $k\in \Zbb_{\geq 0}/2$, the module $W_k$ is simple, and hence $I_k=W_k$.
  For the $\TL^a$ standard modules, there are two subcases. For any $k\in \Zbb_{\geq 0}/2$, if $z$ is of the form $z=\eps q^{\sigma m}$ with $\eps,\sigma \in \{-1,+1\}$ and $m\in k+\Zbb_{>0}$, then one has $W_{k,\eps q^{\sigma m}} \equiv [I_{k,\eps q^{\sigma m}}\to I_{m,\eps q^{\sigma k}}]$. Otherwise, $W_{k,z}$ is simple, and hence $I_{k,z}=W_{k,z}$.
\end{enumerate}

\paragraph{Braid transfer matrices.}
Let $y\in \Cbb^\times$ such that $y^2=q$ or $y^2=q^{-1}$.
The operators $b_j,\bb_j$ defined as
\begin{equation}
  b_j := y\, \id + y^{-1}\, e_j \,,
  \qquad \bb_j := y^{-1}\, \id + y\, e_j 
\end{equation}
satisfy the braid relations
\begin{equation}
  b_j b_{j+1} b_j = b_{j+1} b_j b_{j+1} \,,
  \qquad \bb_j \bb_{j+1} \bb_j = \bb_{j+1} \bb_j \bb_{j+1} \,,
  \qquad b_j \bb_j = \bb_j b_j = \id \,.
\end{equation}
The braid transfer matrices for $\TL(N)$ and $\TL^a(N)$ read, respectively
\begin{equation} \label{eq:G}
  G = c_1 b_2 b_3\dots b_{N+1}b_{N+1} \dots b_3b_2 c^\dag_1
  = c_1 \bb_2 \bb_3\dots \bb_{N+1}\bb_{N+1} \dots \bb_3\bb_2 c^\dag_1 \,,
\end{equation}
and
\begin{equation} \label{eq:F}
  F = c_{N+1} \bb_N \bb_{N-1}\dots \bb_1 \, c^\dag_0 \,,
  \qquad \Fb = c_{N+1} b_N b_{N-1}\dots b_1 \, c^\dag_0 \,,
\end{equation}
The operators $G$ and $F,\Fb$ are central elements of $\TL(N)$ and $\TL^a(N)$ respectively, and they act in simple modules as
\begin{equation} \label{eq:F.I} 
  G \big|_{I_k(N)} = (-q^{2k+1}-q^{-2k-1}) \,\id \,,
  \qquad \begin{cases}
    F \big|_{I_{k,z}(N)} = (y^{2k}z + y^{-2k}z^{-1})\,\id \,, \\
    \Fb\big|_{I_{k,z}(N)} = (y^{2k}z^{-1} + y^{-2k}z)\,\id \,.
  \end{cases}
\end{equation}

\subsection{Scaling limit}

The scaling limit considered in this paper consists in setting all the coupling constants to $v=1$ in \eqref{eq:face}, and computing partition functions on an $N\times M$ lattice in the limit $M,N\to\infty$. Equivalently, this limit is encoded in the low-energy spectrum of the Hamiltonian
$$ H = -\sum_{j=1}^N e_j $$
as $N\to \infty$.
The following results derive from the isomorphism \cite{BKW76} between the standard module $W_{k,w}$ and the vector space with magnetisation $S^z=k$ in the XXZ spin chain with anisotropy parameter $\Delta=\beta/2$, and diagonal twisted boundary conditions $(\sigma^x\pm i\sigma^y) \to w^{\pm 1} (\sigma^x\pm i\sigma^y)$ and $\sigma^z \to \sigma^z$.
\medskip

If $q$ is a root of unity, we write $q=e^{i\pi h'/h}$ as above.
The related XXZ spin chain has $\Delta=\cos(\pi h'/h)$ and thus it is in the critical phase.
The scaling limit of the simple $\TL^a$-modules is then described in terms of CFT, with the Kac parameterisation of the central charge and conformal weights
\begin{equation}
  c = 1-\frac{6(h-h')^2}{hh'} \,,
  \qquad \Delta_{rs} = \frac{(rh-sh')^2 - (h-h')^2}{4hh'} \,.
\end{equation} Let $s,\sbar$ be two integers such that $1\leq s,\sbar \leq h-1$.  As $N\to \infty$, the low-energy part of  $\I_{s,\sbar}(N)$~\eqref{eq:I} scales to
\begin{equation} \label{eq:scaling.I}
  \I_{s,\sbar}(N) \to \bigoplus_{r=1}^{h'-1} [\Delta_{rs},\Delta_{r\sbar}] \,,
\end{equation}
where $[\Delta,\bar\Delta]$ denotes the irreducible $(\mathrm{Vir}\otimes \overline{\rm Vir})$ module with heighest weights $\Delta$ and $\bar\Delta$, and $(\mathrm{Vir},\overline{\rm Vir})$ are two copies of the Virasoro algebra of central charge $c$.
Moreover, on the states of $\I_{s,\sbar}$ which scale to $[\Delta_{rs},\Delta_{r\sbar}]$, the transfer matrix and shift operator act as
\begin{equation} \label{eq:scaling.T.Om}
  t_N \to (-1)^{r+s} \, \exp\left[-\frac{2\pi}{N}(L_0+\Lb_0-\frac{c}{12})\right] \,,
  \qquad \Omega \to (-1)^{r+s} \, \exp\left[-\frac{2i\pi}{N}(L_0-\Lb_0)\right] \,.
\end{equation}
\medskip

The other case which is considered in the present work is when the loop weight $\beta$ is greater than or equal to $2$. The related XXZ spin chain has $\Delta>1$ and thus it is in the gapped phase. The face model is not critical, and the correlation length remains finite in the scaling limit.
\medskip

\subsection{Seed states}

Let $A$ be a $\TL$-module. A seed state $\xi$ of type $k$ in $A$ is a nonzero state $\xi\in A(2k)$ such that $c_1\,\xi=\dots=c_{2k-1}\,\xi=0$, i.e. the action of a diagram on $\xi$ is zero if this diagram connects two nodes of $\xi$. In other words, the state $\xi$ behaves like $u_k\in W_k(2k)$ under the action of $\TL(N,2k)$. Then, one can show that the \textit{insertion map}
\begin{equation} \label{eq:phi:Vk->M}
  \varphi:\lambda\cdot u_k \mapsto \lambda\cdot \xi
\end{equation}
is well-defined for all $\lambda\in \TL(N,2k)$, and yields a morphism of modules from $W_k$ to $A$.
\medskip

Similarly, let $A$ be a $\TL^a$-module. A seed state of type $(k,z)$ in $A$ is a non-zero element $\xi\in A(2k)$ such that
\begin{equation}
  \begin{cases}
    c_1 \cdot \xi = 0 & \text{if } k\geq 1 \,, \\
    \Omega\cdot \xi=z\, \xi & \text{if } k\geq 1/2 \,, \\
    f\cdot \xi = (z+z^{-1})\,\xi & \text{if } k=0 \,.
  \end{cases}
\end{equation}
Then the map  \eqref{eq:phi:Vk->M} is well-defined for all $\lambda \in \TL^a(N,2k)$, and yields a morphism of $\TL^a$-modules from $W_{k,z}$ to $A$.
\medskip

In the following, we suppose furthermore that the $\TL$-module (resp. $\TL^a$-module) $A$ is equipped with an inner product $\aver{\,,\,}$ which is invariant, i.e. $\aver{u,\lambda\,v}=\aver{\lambda^\dag\,u,v}$ for any $u \in A(N),\ v \in A(N'),\ \lambda \in \TL(N,N')$ [resp. $\lambda \in \TL^a(N,N')$]. We then prove three properties of seed states.
We present only the proofs for $\TL$ modules, because the ones for $\TL^a$ modules are perfectly similar.
\medskip

\noindent\textbf{Property 1.}
Suppose that the loop weight $\beta$ is real.
Let $A$ be a $\TL$-module equipped with an invariant inner product $\aver{\,,\,}$ and $\xi$ be seed state of type $k$ in $A$. Then one has
\begin{equation} \label{eq:<>=Gram}
  \aver{\varphi(u^*),\varphi(v)} = \aver{\xi,\xi}\, \aaver{u,v} \,,
\end{equation}
for any $u,v \in W_k(N)$, where $u^*$ denotes the complex conjugate of $u$ in the link state basis of $W_k(N)$, whereas $\varphi:W_k\to A$ is the insertion map associated to $\xi$, and $\aaver{\,,\,}$ denotes the Gram form in $W_k$.

Similarly, let $A$ be a $\TL^a$ module equipped with an invariant inner product $\aver{\,,\,}$ and $\xi$ be seed state of type $(k,z)$ in $A$, with $|z|=1$. If $\beta$ is real, then \eqref{eq:<>=Gram} holds for any $u\in W_{k,z^*}(N)$ and $v \in W_{k,z}(N)$. In this case, $u^*$ is considered as an element of $W_{k,z}(N)$.
\medskip

\noindent\textbf{Proof.} For any $\lambda \in \TL(N,N')$, denote by $\lambda^*$ the complex conjugate of $\lambda$ in the basis of diagrams.
Since $\beta$ is real, all the diagrams of $\TL(N,N')$ have real matrix elements in the link state bases of $W_k(N),W_k(N')$, and we thus have
 $\lambda^*\cdot u^* = (\lambda\cdot u)^*$ for any $\lambda \in\TL(N,N')$. We introduce the bilinear form $f:W_k(N) \times W_k(N) \to \Cbb$ defined as $f(u,v)=\aver{\varphi(u^*),\varphi(v)}$, where $\varphi$ is the morphism \eqref{eq:phi:Vk->M} associated to $\xi$. For any $u\in W_k(N), \ v\in W_k(N')$ and $\lambda \in \TL(N,N')$, one has
\begin{equation*}
  f(u,\lambda\,v)
  = \aver{\varphi(u^*),\varphi(\lambda\,v)}
  = \aver{\varphi(\lambda^\dag\,u^*),\varphi(v)}
  = \aver{\varphi((\lambda^t\,u)^*),\varphi(v)} = f(\lambda^t\,u, v) \,,
\end{equation*}
and thus $f$ is an invariant bilinear form.
On the other hand, we can write $f(u_k,u_k)=\aver{\xi,\xi}$. Hence, using the uniqueness of the Gram form,  we obtain \eqref{eq:<>=Gram}.
\eproof
\medskip

\noindent\textbf{Property 2.} Under the conditions of Property 1, one has
\begin{equation} \label{eq:TL.xi=Ik}
  \TL(N,2k) \cdot \xi \equiv I_k(N)
  \qquad \text{and} \qquad \TL^a(N,2k) \cdot \xi \equiv I_{k,z}(N) \,,
\end{equation}
respectively for a seed state $\xi$ of type $k$ and $(k,z)$ in $A$.
\medskip

\noindent\textbf{Proof.} The submodule $L_\xi(N):=\TL(N,2k) \cdot \xi$ is the image of the insertion map:
$$L_\xi(N)=\varphi(W_k(N)) \,.$$
Since $\varphi$ is a morphism, we have $L_\xi \equiv W_k/\mathrm{Ker}\,\varphi$. Let $v \in \mathrm{rad}(W_k(N))$. For any $u'\in L_\xi(N)$, we can write $u'=\varphi(u)$ with $u\in W_k(N)$, which yields
$$\aver{u',\varphi(v)}=\aver{\varphi(u),\varphi(v)}=\aver{\xi,\xi} \aaver{u^*,v}=0 \,,$$
and hence $\varphi(v)=0$. This shows that $\mathrm{rad}(W_k) \subseteq \mathrm{Ker}\,\varphi$. Moreover, since $\varphi(u_k)=\xi\neq 0$, we have $\mathrm{Ker}\,\varphi \neq W_k$. Hence, we obtain $\mathrm{Ker}\,\varphi=\mathrm{rad}(W_k)$, which proves \eqref{eq:TL.xi=Ik}.
\eproof
\medskip

\noindent\textbf{Property 3.}
Let $A$ be a $\TL$-module equipped with an invariant inner product $\aver{\,,\,}$, and $\xi,\xi'$ be two seed states of type $k,k'$, respectively, with $\aver{\xi,\xi'}=0$ in the case $k=k'$.
Then the modules $\TL(N,2k) \cdot \xi$ and $\TL(N,2k) \cdot \xi'$ are orthogonal.
Similarly, the property holds if $A$ is a $\TL^a$ module and $\xi,\xi'$ are seed states of types $(k,z),(k',z')$, respectively, with $\aver{\xi,\xi'}=0$ in the case $k=k'$.
\medskip

\noindent\textbf{Proof.} Let us write $u=\lambda\,\xi$ and $v=\mu\,\xi'$, where $\lambda$ is a diagram of $\TL(N,2k)$, and $\mu$ is a diagram of $\TL(N,2k')$. Then
\begin{equation}
  \aver{\lambda\,\xi, \mu\,\xi'} = \aver{\mu^\dag\lambda\,\xi, \xi'} = \aver{\xi, \lambda^\dag\mu\,\xi'} \,.
\end{equation}
If $k>k'$, we have $\mu^\dag\lambda\in\TL(2k',2k)$, and hence $\mu^\dag\lambda\,\xi=0$. Similarly, if $k<k'$, we obtain $\lambda^\dag\mu \,\xi'=0$. If $k=k'$, we have $\mu^\dag\lambda\,\xi=0$, unless $\mu^\dag\lambda$ is propotional to the diagram $\id$. In every case, we obtain $\aver{\lambda\,\xi, \mu\,\xi'} = 0$.
\eproof

\section{Chebyshev polynomials and binomial formulas}
\label{app:cheby}

\paragraph{Chebyshev polynomials.}
The Chebyshev polynomial of the first and second kind $T_m$ and $U_m$ are defined by the relations
$$T_m(\cos\theta) = \cos m\theta \,, \qquad U_m(\cos\theta) = [\sin(m+1)\theta]/\sin\theta$$
or, equivalently, by the recursion formulas
\begin{equation}
  \begin{aligned}
    &T_0(x) =1 \,, &\qquad& T_1(x) = x \,,
    &\qquad& T_{m+1}(x) = 2x \,T_m(x) - T_{m-1}(x) \,, \\
    & U_0(x) =1 \,, && U_1(x) = 2x \,, && U_{m+1}(x) = 2x \,U_m(x) - U_{m-1}(x) \,.
  \end{aligned}
\end{equation}
Moreover, they obey the property:
\begin{equation} \label{eq:diff.Um}
  U_m(x)-U_{m-2}(x) = 2T_m(x) \,.
\end{equation}

\paragraph{Binomial formulas.}
Using the Newton binomial formula, one easily derives the identity:
\begin{equation} \label{eq:Newton}
  (2\cos\theta)^N= \sum_{k \in \Zbb/2} d_k(N) \,\cos 2k\theta \,,
\end{equation}
where $d_k(N)$ is defined in \eqref{eq:def.dk}.
For any non-zero integers $h,m$ we obtain from the above relation
\begin{equation} \label{eq:Newton2}
  \left(2\cos \frac{\pi m}{h} \right)^N= \sum_{k=0,\frac12,\dots,h-\frac12} D_k(N) \,\cos \frac{2\pi km}{h} \,,
\end{equation}
where $D_k(N)$ is defined in \eqref{eq:def.Dk}.


\begin{thebibliography}{99}

\bibitem{PZ01} V.B. Petkova and J.-B. Zuber,
  \textit{Generalised twisted partition functions},
  Phys. Lett. {\bf B 504}, 157--164 (2001)

\bibitem{Manyfaces01} V.B. Petkova and J.-B. Zuber,
  \textit{The many faces of Ocneanu cells},
  Nucl. Phys. {\bf B 603}, 449--496 (2001)

\bibitem{Ocneanu99} A. Ocneanu,
  \textit{Paths on Coxeter diagrams: From Platonic solids and singularities to minimal models and subfactors},
  in Lectures on Operator Theory, Fields Institute, Waterloo, Ontario, April 26–30, 1995, (Notes taken by S. Goto) Fields Institute Monographies, AMS 1999, Rajarama Bhat et al, eds.
  
\bibitem{Ocneanu00} A. Ocneanu,
  \textit{The Classification of subgroups of quantum SU(N)},
  Lectures at Bariloche Summer School 2000, Argentina, AMS Contemp. Math. 294, R. Coquereaux, A. Garc\'{\i}a and R. Trinchero eds.
  
\bibitem{Chui01} C.H.O. Chui, C. Mercat, W.P. Orrick and P.A. Pearce,
  \textit{Integrable lattice realizations of conformal twisted boundary conditions},
  Phys Lett. {\bf B 517}, 429--435 (2001)

\bibitem{Chui03} C.H.O. Chui, C. Mercat and P.A. Pearce,
  \textit{Integrable and conformal twisted boundary conditions for $sl$(2) A-D-E lattice models},
  J. Phys. A: Math. Gen. {\bf 36}, 2623--2662 (2003)

\bibitem{Frohlich04} J. Fr\"ohlich, J. Fuchs, I. Runkel and C. Schweigert,
  \textit{Kramers-Wannier duality from conformal defects},
  Phys. Rev. Lett. {\bf 93}, 070601 (2004)

\bibitem{Frohlich07} J. Fr\"ohlich, J. Fuchs, I. Runkel and C. Schweigert,
  \textit{Duality and defects in rational conformal field theory},
  Nucl. Phys. {\bf B 763}, 354--430 (2007)

\bibitem{Kitaev06} A. Kitaev, \textit{Anyons in an exactly solved model and beyond},
  Annals of Physics {\bf 321}, 2--111 (2006)

\bibitem{Feiguin06} A. Feiguin, S. Trebst, A.W.W. Ludwig et al,
  \textit{Interacting anyons in topological quantum liquids: The golden chain},
  Phys. Rev. Lett. {\bf 98}, 160409 (2007)

\bibitem{Aasen16} D. Aasen, R.S.K. Mong and P. Fendley,
  \textit{Topological defects on the lattice: I. The Ising model},
  J. Phys. A: Math. Theor. {\bf 49}, 354001 (2016)
  
\bibitem{Aasen20} D. Aasen, P. Fendley and R.S.K. Mong,
  \textit{Topological Defects on the Lattice: Dualities and Degeneracies},
  arXiv:2008.08598v1 [cond-mat.stat-mech]

\bibitem{Lootens23} L. Lootens, C. Delcamp, G. Ortiz and F. Verstraete,
  \textit{Dualities in one-dimensional quantum lattice models: symmetric Hamiltonians and matrix product operator intertwiners},
  PRX Quantum {\bf 4}, 020357 (2023)

\bibitem{Lootens24} L. Lootens, C. Delcamp and F. Verstraete,
  \textit{Dualities in one-dimensional quantum lattice models: topological sectors},
  PRX Quantum {\bf 5}, 010338 (2024)

\bibitem{Bottini26} L. Bhardwaj, L.E. Bottini, S. Schafer-Nameki and A. Tiwari,
  \textit{Lattice Models for Phases and Transitions with Non-Invertible Symmetries},
  SciPost Phys. {\bf 20}, 134 (2026)

\bibitem{TL71} H.~Temperley and E.~Lieb,
  \textit{Relations between the ``percolation'' and ``colouring'' problem and
    other graph-theoretical problems associated with regular planar lattices:
    Some exact results for the ``percolation'' problem},
  Proc. Roy. Soc. London Ser. {\bf A322}, 251--280 (1971)

\bibitem{ABF84} G.E.~Andrews and R.J.~Baxter and P.J.~Forrester,
  \textit{Eight-vertex SOS model and generalized Rogers--Ramanujan-type identities},
  J. Stat. Phys. {\bf 35}, 193--266 (1984)

\bibitem{PasquierADE87} V.~Pasquier,
  \textit{Two-dimensional critical systems labelled by Dynkin diagrams},
  Nucl. Phys. {\bf B 285}, 162--172 (1987)

\bibitem{PasquierOpContent87} V.~Pasquier,
  \textit{Operator content of the ADE lattice models},
  J. Phys. A: Math Gen. 20, 5707--5717 (1987)


\bibitem{Belletete20} J. Bellet\^ete, A.M. Gainutdinov, J.L. Jacobsen et al,
  \textit{Topological defects in periodic RSOS models and anyonic chains},
  arXiv:2003.11293v1 [math-ph]
  
\bibitem{Grans24} M. Sinha, F. Yan, L. Grans-Samuelsson et al.
  \textit{Lattice realizations of topological defects in the critical (1+1)-d three-state Potts model},
  J. High Energ. Phys. 2024, 225 (2024)
  
\bibitem{GL98} J.J.~Graham and G.I.~Lehrer,
  \textit{The representation theory of affine Temperley--Lieb algebras},
  Ens.~Math. {\bf 44}, 173--218 (1998)

\bibitem{IMD-RSOS} Y.~Ikhlef and A.~Morin-Duchesne,
  \textit{Temperley--Lieb modules and local operators for critical ADE models},
  arXiv:2602.15742 [math-ph]
      
\bibitem{BPZ98}
  R.E.~Behrend, P.A.~Pearce and J.-B.~Zuber,
  \textit{Integrable Boundaries, Conformal Boundary Conditions and A-D-E Fusion Rules},
  J. Phys. {\bf A 31}, L763--L770 (1998)

\bibitem{BPPZ98}  R.E. Behrend, P.A. Pearce, V.B. Petkova and J.-B. Zuber,
  \textit{On the classification of bulk and boundary conformal field theories},
  Phys. Lett. {\bf B 444}, 163--166 (1998)

\bibitem{Blakeney26} M. Blakeney, L. Corcoran, M. de Leeuw et al,
  \textit{Temperley--Lieb integrable models and fusion categories},
  J. High Energ. Phys. 2026, 165 (2026)

\bibitem{Verstraete17} D.J. Williamson, N. Bultinck and F. Verstraete,
  \textit{Symmetry-enriched topological order in tensor networks: Defects, gauging and anyon condensation},
  arXiv:1711.07982 [quant-ph]
  
\bibitem{KL94} L.H. Kauffman and S.L. Lins,
  \textit{Temperley--Lieb recoupling theory and invariants of 3-manifolds},
  Annals of Mathematics Studies, Vol. 134, MR1280463, Princeton, NJ: Princeton
  University Press, 1994, pp. x+296.

\bibitem{Guide} C. Edie-Michell and S. Morrison,
  \textit{A field guide to categories with $A_n$ fusion rules},
  arXiv:1710.07362 [math.QA]
  
\bibitem{TY98} D. Tambara and S. Yamagami,
  \textit{Tensor Categories with Fusion Rules of Self-Duality for Finite Abelian Groups},
  Journal of Algebra {\bf 209}, 692--707 (1998)

\bibitem{PasquierSaleur90} V. Pasquier and H. Saleur,
  \textit{Common structures between finite systems and conformal field theories through quantum groups},
  Nucl. Phys. {\bf B330}, 523--556 (1990)

\bibitem{PTLFunctor} Y.~Ikhlef and A.~Morin-Duchesne,
  \textit{Fusion in the periodic Temperley--Lieb algebra: general definition of a bifunctor},
  arXiv:2509.11756 [math-ph]
  
\bibitem{PinetStAubin} Y. Saint-Aubin and T. Pinet,
  \textit{Spin chains as modules over the affine Temperley--Lieb algebra},
  Alg. Repr. Th. {\bf 26}, 2523--2584 (2023)
  
\bibitem{RSA14} D.~Ridout and Y.~Saint-Aubin,
  \textit{Standard modules, induction and the structure of the Temperley--Lieb algebra},
  Adv.~Theor.~Math.~Phys. {\bf 18}, 957--1041 (2014)

\bibitem{BKW76} R.J. Baxter, S.B. Kelland and F.Y. Wu,
  \textit{Equivalence of the Potts model or Whitney polynomial with an ice-type model},
  J. Phys. A: Math. Gen. {\bf 9}, 397 (1976)

\bibitem{CIZ87}
  A.~Cappelli and C.~Itzykson and J.-B.~Zuber,
  \textit{The A-D-E classification of minimal and $A^{(1)}_1$ conformal invariant theories},
  Comm. Math. Phys.{\bf 113}, 1 (1987)

\bibitem{CardyBCFT89} J.L.~Cardy,
  \textit{Boundary conditions, fusion rules and the Verlinde formula},
  Nucl. Phys. {\bf B 324}, 581--596 (1987)
  
\end{thebibliography}
\end{document}